\documentclass{article}
\usepackage{pdfpages}

\usepackage[ansinew]{inputenc}
\usepackage{graphicx}
\usepackage{color}
\usepackage[active,new,noold,marker]{xrcs}

\usepackage{CJK}
\usepackage{graphics}
\usepackage{upgreek}
\usepackage{multicol}
\usepackage{amsmath}
\usepackage[varg]{txfonts} % Times fonts
\usepackage{bm}

\usepackage{ulem}
\usepackage[T1]{fontenc}
\usepackage{calligra}

\usepackage{graphicx}% Include figure files
\usepackage{dcolumn}% Align table columns on decimal point
\usepackage{bm}% bold math

\usepackage{times}
\usepackage{verbatim}
\usepackage{graphicx}
\usepackage{graphics,epsfig}

\usepackage{theorem}
\usepackage{makeidx}
\usepackage{amsmath}
\usepackage{epic}
\usepackage{amscd}
\usepackage{bm}

\usepackage{bbm}
\usepackage{xy}
\usepackage{amssymb}

\usepackage{theorem}
\usepackage{makeidx}
\usepackage{amsmath}
\usepackage{epic}
\usepackage{amscd}

\usepackage{bbm}
\usepackage{xy}
\usepackage{amssymb}
\usepackage{epsfig}

\usepackage{cancel}

\catcode`\=13
\def{$\bowtie$}

\usepackage{eurosym}
\DeclareInputText{128}{\euro} % ANSI code for euro: ? \usepackage{eurosym}
\DeclareInputText{165}{\yen}  % ANSI code for yen:  ? \usepackage{amssymb}

\usepackage{lscape} %landcape pages support
\title{%Conjecturing nonuniversal spectral transfer directions of 2D incompressible passive scalar energy and examining the geophysical fluid dynamics relevances
2D2C, 2D3C, 2C2Dcw1C3D, 3D3C, rotating turbulence, thin-layer flows, quasi-static magnetohydrodynamics (QSMHD), and beyond}
\author{Jian-Zhou Zhu\\
Su-Cheng Centre for Fundamental and Interdisciplinary Sciences, Gaochun, Nanjing, 211316% China
, and, \\
Life and Chinese Medicine Study Center, Gui-Lin Tang Lab., Yong'an, 366025 Fujian, China}
\date{}
\begin{document}
\maketitle
%\centerline{\it In Proceedings of Workshop}
%\centerline{\it ``Turbulence Modeling and Vortex Dynamics''}
%\centerline{\it Istanbul, Turkey,}
%\centerline{\it Springer Lect. Notes in Phys. {\bf 491}, 53--64, 1997.}
%----------------------------------------------

\def\ind{{_E}}
\def\EV{\nu_\ind}
\begin{abstract}
More serious works on 2D2C, 2D3C, 2C2Dcw1C3D, 3D3C, rotating turbulence,
thin-layer flows, quasi-static magnetohydrodynamics (QSMHD), and all that are wanted,
but we report timely here some studies on locally and globally 2C2Dcw1C3D flows, with
the hope to promote smarter and deeper works.
\end{abstract}
%---------------------------------------------------------

\section*{\textbf{0} Preamble}
%\section{Introduction}


# Vorticity and helicity decompositions and dynamics with real Schur form of the velocity gradient

Jian-Zhou Zhu[1, a)]

*Su-Cheng Centre for Fundamental and Interdisciplinary Sciences, Gaochun, Nanjing, China*

The real Schur form (RSF) of a generic velocity gradient field $\nabla \boldsymbol{u}$ is exploited to expose the structures of flows, in particular our field decomposition resulting in two vorticities with only mutual linkage as the topological content of the global helicity (accordingly decomposed into two equal parts). The local transformation to RSF may indicate alternative (co)rotating frame(s) for specifying the objective argument(s) of the constitutive equation. When $\nabla \boldsymbol{u}$ is uniformly of RSF in a fixed Cartesian coordinate frame, i.e., $u_x = u_x(x,y)$ and $u_y = u_y(x,y)$, but $u_z = u_z(x,y,z)$, the model, with the decomposed vorticities both frozen-in to $\boldsymbol{u}$, is for two-component-two-dimensional-coupled-with-one-component-three-dimensional flows inbetween two-dimensional-three-component (2D3C) and fully three-dimensional-three-component ones and may help curing the pathology in the helical 2D3C absolute equilibrium, making the latter effectively work in more realistic situations.

A gradient with respect to $\boldsymbol{r}$, of velocity $\boldsymbol{u} = (u_x, u_y, u_z)$ in $\boldsymbol{r} = (x,y,z)$, can be decomposed into the symmetric ($D = \frac{G+G^T}{2}$) and asymmetric parts ($A = \frac{G-G^T}{2}$),

$$G = \nabla_{\boldsymbol{r}} \boldsymbol{u} = \begin{pmatrix} u_{x,x} & u_{x,y} & u_{x,z} \\ u_{y,x} & u_{y,y} & u_{y,z} \\ u_{z,x} & u_{z,y} & u_{z,z} \end{pmatrix} = D + A, \quad (1)$$

where the index, of each component, before a comma denotes the coordinate component and that following a comma denotes the partial derivative $\partial$ with respect to it. $D$ is conventionally called the strain rate tensor describing the rate of change of the deformation, and $A$ corresponds to the conventional 'vorticity' describing the rigid rotation. The strain rate tensor is composed of the compression/expansion rate tensor and the shear rate tensor, $D = C + S$, with $C = Tr(D)\delta_{ij}$. However, a single element of $G$ only tells the deformaton in a certain direction but not the bulk behavior of the fluid parcle; so, more precisely, all the decomposed components (of $C$, $S$ and $A$) are the algebraic mean rates over directions of the expansion/compression, shear and rotation, respectively. So, further decompositions or more specific quantifications may be useful. For example, relevant to the issue of 'objectivity' in continuum mechanics (e.g., Wedgewood[1], and references therein) there have been many intersting and meaningful propositions of vorticity decompositions. Definitely, any mathmatically consistent decompositions may be of potential theoretical value, and we will propose a natural decomposition based on the real Schur form (RSF)[2] of the velocity gradient.

We consider the velocity gradient $u_{i,j}$ for $1 \to x$, $2 \to y$ and $3 \to z$ in the Euclidean 3-space: Moffatt and Tsinober[3], in their Fig. 2 with three mutually linked vorticity lines, have decomposed the vorticity into three parts, $\boldsymbol{\omega} = \boldsymbol{\omega}_1 + \boldsymbol{\omega}_2 + \boldsymbol{\omega}_3$, with lines of the horizontal vorticity

$$\boldsymbol{\omega}_3 = \nabla \times \boldsymbol{u}_3 = (u_{z,y}, -u_{z,x}, 0) \quad (2)$$

[the index with respect to which the gradient is taken is omitted when it is self-evident] given by

$$u_z = constant, z = constant, \quad (3)$$

and similarly the other two indexed by $y$ and $z$. With $\boldsymbol{u}_1 = (u_x, 0, 0)$, $\boldsymbol{\omega}_1 \cdot \boldsymbol{u}_1 = 0$, and similarly for the other two components, the helicity is then accordingly decomposed into the contributions from the mutual linkages of three types of vorticity lines. However, such a decomposition does not keep, for $\boldsymbol{\omega}_{1,2,3}$, the original vorticity *frozen-in* transport property,[4] i.e., $\partial_t \boldsymbol{\omega} = \nabla \times (\boldsymbol{u} \times \boldsymbol{\omega})$ or any variants of it, losing some dynamical significance. For example, one can check that $\boldsymbol{\omega}_1$ is neither frozen-in to $\boldsymbol{u}$, nor to $\boldsymbol{u}_1$. It is thus an intriguing problem whether there exists a decomposition of the vorticity which results in simple helicity decomposition and decomposed vorticity transport in a fashion as the original total vorticity. This turns out possible in a slightly reduced situation as will be shown below.

A third motivation is relevant to the phenomenological identification of 'vortex structures'.[5] We remark that we also offer definite decomposition of the velocity, not only the vorticity, unlike most others.

When $u_{z,x} = u_{z,y} = 0$, $G$ is of RSF, i.e., the (block/quasi-)upper triangular matrix form from the standard *real Schur transformation/decomposition* (RST/D) of $\nabla_{\boldsymbol{R}} \boldsymbol{U}$ of a *generic* velocity $\boldsymbol{U} = (U_X, U_Y, U_Z)$ in $\boldsymbol{R} = (X,Y,Z)$, and will have a 2-by-2 sub-matrix, arranged here to be the upper-left element, containing conjugate pairs of complex eigenvalues, if exist. The $z$ axis is of special role. So, we write

$$u_i = u_i^{(1)} + u_i^{(2)} : \boldsymbol{u}^{(1)} = (u_x, u_y, 0), \boldsymbol{u}^{(2)} = (0,0,u_z). \quad (4)$$

We then have $\boldsymbol{\omega}^{(1)} = (-u_{y,z}, u_{x,z}, u_{y,x} - u_{x,y})$ and $\boldsymbol{\omega}^{(2)} = \boldsymbol{0}$. We immediately see that the helicity is contributed only by the self-linkage of the vorticity $\boldsymbol{\omega}^{(1)}$ of the horizontal flow $\boldsymbol{u}^{(1)}$. [Pointwise, $\boldsymbol{u}^{(2)} \cdot \boldsymbol{\omega}^{(1)}$ does not vanish, but the integral over the volume element $\text{dvol}$ (here $= dxdydz$) of the domain results in null helicity component $H^{1,2} = \int \boldsymbol{u}^{(2)} \cdot \boldsymbol{\omega}^{(1)} d\text{vol}$, as can be seen by integration by parts.]

RST is realized with a space-time dependent matrix $Q$ for rotation from the $XYZ$ coordinates to $xyz$, dynamically up to a solid body rotation, i.e., the anti-symmetric $Q\dot{Q}^T$ which

---

a)Electronic mail: jz@sccfis.org

does not exist when just the snapshot is considered or vanishes in a steady flow with $\dot{Q} = 0$, say, or, otherwise, which can be eliminated by *locally*[6] rotating the $xyz$ frame [c.f., e.g. Eqs. (8, 24 and 32) of Wedgewood.[1]] The specification of the rotations should refer to the dynamics of the velocity gradient, the gradient of the momentum equation, which is not pursued here.

When RSF holds uniformly in a fixed Cartesian frame, the original vorticity frozen-in property of an ideal flow

$$\partial_t(\boldsymbol{\omega}^{(1)} + \boldsymbol{\omega}^{(2)}) = \nabla \times [(\boldsymbol{u}^{(1)} + \boldsymbol{u}^{(2)}) \times (\boldsymbol{\omega}^{(1)} + \boldsymbol{\omega}^{(2)})] \quad (5)$$

becomes $\partial_t \boldsymbol{\omega}^{(1)} = \nabla \times [\boldsymbol{u} \times \boldsymbol{\omega}^{(1)}]$. In words, the three-dimensional (3D) total vorticity $\boldsymbol{\omega} = \boldsymbol{\omega}^{(1)}$ is only due to the horizontal 3D-two-component (2C) flow $\boldsymbol{u}^{(1)}$ and is frozen-in to the total flow.

It appears that we do not get much in the above, but note that if we perform the RST for the transposition of the original gradient matrix or switch the $x$ and $z$ axises, we have

$$u_{x,z} = 0 = u_{y,z}, \quad (6)$$

with which locally two-component-two-dimensional-coupled-with-one-component-three-dimensional (2C2Dcw1C3D) flow results, and with which we here also call $G$ a RSF. Then, from Eq. (4), we now have

$$\boldsymbol{\omega}^{(1)} = (0, 0, u_{y,x} - u_{x,y}), (7)$$

and, as in Eqs. (2 and 3), $\boldsymbol{\omega}^{(2)} = (u_{z,y}, -u_{z,x}, 0) = \boldsymbol{\omega}_3.(8)$

Two important facts follow:

(1) $\boldsymbol{u}^{(1)} \cdot \boldsymbol{\omega}^{(1)} = \boldsymbol{u}^{(2)} \cdot \boldsymbol{\omega}^{(2)} = 0$, which leads to the helicity decomposition

$$H = H^{1,2} + H^{2,1} = \int_{\mathcal{D}} (\boldsymbol{\omega}^{(1)} \cdot \boldsymbol{u}^{(2)} + \boldsymbol{\omega}^{(2)} \cdot \boldsymbol{u}^{(1)}) d\texttt{vol}, \quad (9)$$

corresponding to the local helicity density decomposition (in the above integrand). The equality $H^{1,2} = H^{2,1}$ holds, if integration by parts contributes nothing from the boundaries, meaning that the global helicity is accordingly decomposed into equal parts. In terms of topological interpretation, the self-linkages of vorticity lines are zero, leaving the mutual-linkage between $\boldsymbol{\omega}^{(1)}$ and $\boldsymbol{\omega}^{(2)}$: A sketch in Fig. 1 is for the extra global 2C2Dcw1C3D condition of uniform $u_{x,z} = 0 = u_{y,z}$ in fixed Cartesian coordinates for the domain $\mathcal{D}$ (the additional incompressibility condition can tell more information of $u_z$ as we will see below). Note that straight vertical vorticity lines should extend to the boundary (see, e.g., Cromwell[10] for a general discussion of working in $R^3$ or the 3-sphere $S^3$ concerning the point $\infty$), and for horizontal vorticity lines, loops can only be trivial knots in the plane.

(2) $\nabla \times \boldsymbol{u}^{(2)} \times \boldsymbol{\omega}^{(2)} = \boldsymbol{0} = \boldsymbol{u}^{(2)} \times \boldsymbol{\omega}^{(1)}$ for RSF holding uniformly in a fixed Cartesian coordinate frame, which by Eq. (5) and some more algebra (as partly given explicitly below as a proof before discussing the relaxation dynamics, both vorticities are frozen-in to $\boldsymbol{u}$

$$\partial_t \boldsymbol{\omega}^{(i)} = \nabla \times [\boldsymbol{u} \times \boldsymbol{\omega}^{(i)}] : i = 1, 2. \quad (10)$$

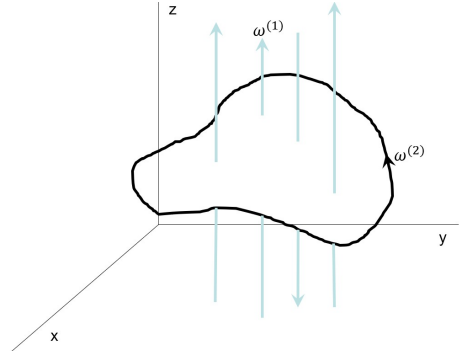

FIG. 1. Mutual linkage between the closed horizontal vorticity lines and the straight lines of vertical vorticity.

With such a formulation in the fashion of Helmholtz, many classical ideal vorticity dynamics carry over to these two vorticities.

We note that whether or not $u_{z,z}$ vanishes does not change the above results [except for an even stronger result that $\boldsymbol{\omega}^{(2)}$ is frozen-in also to $\boldsymbol{u}^{(1)}$ with $u_{z,z} = 0$]. In other words, the above results hold also for 2D3C situation which is marked in rapidly rotating flows (ubiquitous in atmosphere of astrophysical objects) as vortical modes or in purely 2D passive scalars. [Note that Moffatt[7], in his Sec. 6. (a), already addressed the 2D3C situation, i.e., with further $u_{z,z} = 0$, but the helicity decomposition and individual vorticity transport issues were not discussed there.] However, with $u_{z,z} \neq 0$, one important difference compared to the 2D3C situation is that the vertical velocity $u_z$ is no longer passively advected by the horizontal flow $\boldsymbol{u}^{(1)}$ (see more below). Such a flow may be treated as a system for the interactions between the vertically-averaged horizontal velocity and the (unaveraged) vertical one, which may be a particularly good approximation for the resonant wave theory (RWT, see, e.g., Greenspan[8] and references therein) of rotating flow, especially when subjected further to such effects that the horizontal velocity is two-dimensionalized much faster (thus yet another time scale is introduced in the RWT) and/or much severe than the vertical one. The selective stratification of $u_z$ can also be realized in *active fluids* (see, e.g., Marchetti et al.[9]), where the microswimmers or molecular motors, for instance, can help. In the context of RWT, some of the wave/fast-vortex/slow interactions fall into or emerge as such a "minimal" model.

We put down the Navier-Stokes equations with Eq. (6),

$$\partial_t \boldsymbol{u}_h + \boldsymbol{u}_h \cdot \nabla_h \boldsymbol{u}_h = -\nabla_h p + \nu \nabla_h^2 \boldsymbol{u}_h, (11)$$
$$\partial_t u_z + \boldsymbol{u}_h \cdot \nabla_h u_z + u_z u_{z,z} = -p_{,z} + \nu(\nabla_h^2 + \cancel{\partial_{zz}})u_z, (12)$$
$$\nabla_h \cdot \boldsymbol{u}_h + u_{z,z} = 0. (13)$$

In the above, $\nu$ is the kinematic viscosity and the index '$h$' is used to denote the horizontal '$x - y$' plane onto which $\boldsymbol{u}$ or $\boldsymbol{u}^{(1)}$ is projected to be $\boldsymbol{u}_h$. The continuity equation given by the incompressibility (13) shows that $u_z$ feeds back onto $\boldsymbol{u}^{(1)}$,

and viscous damping of $u_z$ is only in the horizontal plane:

$$\partial_z \boldsymbol{u}_h = 0 \text{ in (13)} \Rightarrow u_{z,zz} = 0 \text{ (denoted by a slash)}, \quad (14)$$
$$\text{and, } \partial_z \nabla_h p = 0 \text{ in Eq. (11)}. \quad (15)$$

The viscous linear damping terms will not be included in our nonlinear calculations. We now prove the less familiar part of Eq. (10) for $\boldsymbol{\omega}^{(2)}$:

*Proof.* By using $\nabla(\boldsymbol{a} \cdot \boldsymbol{b}) = \boldsymbol{b} \cdot \nabla \boldsymbol{a} + \boldsymbol{a} \cdot \nabla \boldsymbol{b} + \boldsymbol{a} \times (\nabla \times \boldsymbol{b}) + \boldsymbol{b} \times (\nabla \times \boldsymbol{a})$ twice, respectively $\boldsymbol{a} = \boldsymbol{u}, \boldsymbol{b} = \boldsymbol{u}^{(2)}$ and $\boldsymbol{a} = \boldsymbol{b} = \boldsymbol{u}^{(2)}$, we have:

$$\nabla u_z^2 = \boldsymbol{u}^{(2)} \cdot \nabla \boldsymbol{u} + \boldsymbol{u} \cdot \nabla \boldsymbol{u}^{(2)} + \boldsymbol{u} \times (\nabla \times \boldsymbol{u}^{(2)})$$
$$+ \boldsymbol{u}^{(2)} \times (\nabla \times \boldsymbol{u}), \quad (16)$$
$$\nabla(u_z^2)/2 = \boldsymbol{u}^{(2)} \cdot \nabla \boldsymbol{u}^{(2)} + \boldsymbol{u}^{(2)} \times (\nabla \times \boldsymbol{u}^{(2)}). \quad (17)$$

Comparing the above two equations and using Eqs. (4, 6, 8 and 7), we have

$$\boldsymbol{u}^{(1)} \cdot \nabla \boldsymbol{u}^{(2)} + \boldsymbol{u}^{(1)} \times (\nabla \times \boldsymbol{u}^{(2)}) = 0. \quad (18)$$

Taking Eqs. (18 and 17) into (12) rewritten in terms of $\boldsymbol{u}^{(2)}$, we have what we want:

$$\partial_t \nabla \times \boldsymbol{u}^{(2)} = \nabla \times [\boldsymbol{u} \times (\nabla \times \boldsymbol{u}^{(2)})]. \quad (19)$$
□

We have assumed for Eqs. (11, 12, 13) that Eq. (6) holds uniformly in a fixed Cartesian coordinate frame over some domain $\mathcal{D}$, so, by integrating Eqs. (14, and 15) respectively

$$u_z = z\zeta(x,y,t) + \xi(x,y,t) \quad (20)$$
$$\text{and } p = \Pi(x,y) + \pi(z) \quad (21)$$

the latter of which leads from Eq. (12) to $\partial_t(\nabla_h u_z) + \boldsymbol{u}_h \cdot \nabla_h(\nabla_h u_z) = -\nabla_h(\partial_z u_z^2/2) + \nu \nabla_h^2(\nabla_h u_z)$, thus the same dynamical equations of $\nabla_h \times \boldsymbol{u}_h$ and $\nabla_h \times (\nabla_h u_z)$. On the one hand, it is then indicated that the $u_z$ structure in the $z$ direction is relatively simple, and, on the other hand, the z-boundary conditions are limited to be quite restrictive in such a case. Periodic or vanishing as $z \to \infty$ boundary condition is not possible now. Also, we have applied uniform density $\rho = 1$, but if we let it change in space and time in a barotropic way $[\rho = \rho(p)$, so that the vorticity is still ideally frozen-in to the fluid], richer structures can present.

The 2C2Dcw1C3D system (11,12, 13), is something in between the 2D3C and fully 3D3C Navier-Stokes. Forgetting about Eq. (20) and assuming periodic boundary condition, say, we would have the formal dynamical invariants for the corresponding inviscid Euler equations,

$$2\mathcal{C} = H \text{ in Eq. (9)}, \quad \mathcal{E} = \int_\mathcal{D} |\boldsymbol{u}_h|^2 d\texttt{vol},$$
$$\mathcal{W} = \int_\mathcal{D} |\boldsymbol{\omega}^{(1)}|^2 d\texttt{vol} \text{ and } \mathcal{Z} = \int_\mathcal{D} |u_z|^2 d\texttt{vol},$$

formally the same as the incompressible 2D3C or incompressible passive scalar system with $u_{z,z} = 0$ (and that $\nabla \cdot \boldsymbol{u}_h = 0$)! Now, realizing Eq. (20), we won't be able to have the periodic boundary condition, neither the invariances of $\mathcal{E}$, $\mathcal{W}$ and $\mathcal{Z}$.

We may assume $\zeta$, or $u_{z,z}$, be very small (to be the order of Rossby number in the rotating flow case as in our later remark on the perturbation to the relaxation), to have a system well close to a 2D3C one.

Treating the small $u_{z,z}$ as a perturbation of the 2D3C or two-dimensional passive scalar problem may actually help resolve a pathology in the statistical treatment of the latter for the approximation or the major dynamics responsible for some more complex dynamics. Dynamically speaking, the statistical correlation $\mathcal{C}$ (we will soon see why such a notation is used) between a passive scalar and the (derivative of the) advecting flow is different to that of an active scalar, but the calculation using the correlation as a constraint does not reflect the information concerning 'passive' or 'active'. For instance, following Lee[11] and Kraichnan[12,13], we may perform the corresponding absolute equilibrium calculation to understand or predict some fundamental aspects of 2D3C or 2D passive scalar turbulence. The standard calculation goes as follows: For $\boldsymbol{u}$ in a cyclic box with dimension $2\pi$, we use the Fourier representation $\boldsymbol{u}(\boldsymbol{r}) = \sum_{\boldsymbol{k}} \hat{\boldsymbol{u}}_{\boldsymbol{k}} \exp\{\hat{i}\boldsymbol{k} \cdot \boldsymbol{r}\}$ and introduce the Galerkin truncation $\boldsymbol{u}^g \triangleq \tilde{\sum}_{\boldsymbol{k}} \hat{\boldsymbol{u}}_{\boldsymbol{k}} \exp\{\hat{i}\boldsymbol{k} \cdot \boldsymbol{r}\}$, with $\tilde{\sum}_{\boldsymbol{k}} \triangleq \sum_{k_{min} \leq |\boldsymbol{k}_h| \leq k_{max}}$ and '$\triangleq$' meaning definition. We further introduce the self-evident notations $\boldsymbol{k}_h$ (following those of the physical-space variables — all complex variables are consistently wearing hats in this paper and $\hat{i}$ is the pure imaginary unit). For the incompressible 2D flow, denoting the horizontal wavevector by $\boldsymbol{k}_h$, we have $\hat{\boldsymbol{u}}_h \cdot \boldsymbol{k}_h = 0$ and are left with only one degree of freedom in the direction $\boldsymbol{h}(\boldsymbol{k}_h) = \boldsymbol{z} \times \boldsymbol{k}_h/k_h$ for $\hat{\boldsymbol{u}}_h$; so, writing $\hat{\boldsymbol{u}}_h = \hat{u}_h(\boldsymbol{k}_h)\boldsymbol{h}(\boldsymbol{k}_h)$, we have $\mathcal{C} = \tilde{\sum} \hat{i} k_h (\hat{u}_h \hat{u}_z^* - c.c.)$, $\mathcal{E} = \tilde{\sum} |\hat{u}_h|^2$, $\mathcal{W} = \tilde{\sum} k_h^2 |\hat{u}_h|^2$, and $\mathcal{Z} = \tilde{\sum} |\hat{u}_z|^2$. For simplicity, we have omitted the conventional factors of $1/2$ which can be absorbed into the (inverse) temperature parameters (see below). It can be shown that those dynamical invariants are rugged, in the sense of being still invariant with the Galerkin truncation. It can also be shown that the ordinary dynamical equations in terms of the imaginary and real parts of the truncated Fourier modes satisfy the Liouville theorem, i.e., the flow in the phase space expanded by the variables is incompressible. Then, the system presumably would approach the statistical equilibrium state, under reasonable conditions (usually ergodicity is assumed). Introducing corresponding Lagrange multipliers or the (inverse) temperature parameters $\Gamma_\bullet$, we now apply the canonical ensemble with constant temperature parameters, assigning a probability $\mathcal{P}$ (some thermodynamics limit of which may be called a Gibbs measure) to each microstate

$$\mathcal{P} \sim \exp\{-(\Gamma_\mathcal{C}\mathcal{C} + \Gamma_\mathcal{E}\mathcal{E} + \Gamma_\mathcal{W}\mathcal{W} + \Gamma_\mathcal{Z}\mathcal{Z})\},$$

to obtain the modal spectral densities, for instance, $U_h$ of $\mathcal{E}$ and $U_z$ of $\mathcal{Z}$,

$$U_h \triangleq \langle |\hat{u}_h|^2 \rangle = \frac{\Gamma_\mathcal{Z}}{D}, \quad U_z \triangleq \langle |\hat{u}_z|^2 \rangle = \frac{\Gamma_\mathcal{E} + \Gamma_\mathcal{W} k_h^2}{D}, (22)$$

with $D = \Gamma_\mathcal{E}\Gamma_\mathcal{Z} + (\Gamma_\mathcal{W}\Gamma_\mathcal{Z} - \Gamma_\mathcal{C}^2)k_h^2 > 0$ from the realizability.

We see that the above calculation does not recognize or exploit the information of 'passiveness', and $\Gamma_\mathcal{Z}$ and/or $\Gamma_\mathcal{C} \neq 0$

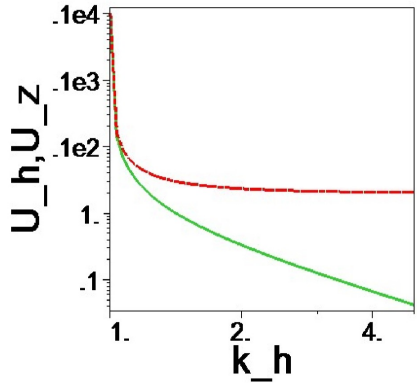

FIG. 2. An example of the modal spectra concentrating at large scales ($k_{min} = 1$): Solid line (blue) for $U_h$ and dash line (red) for $U_z$ are plotted with $\Gamma_{\mathcal{E}} = -1$, $\Gamma_{\mathcal{Z}} = 1$, $\Gamma_{\mathcal{W}} = 2.001$ and $\Gamma_{\mathcal{C}} = -1$.

'wrongly' enter(s) the expression of $U_h$: for a pure passive scalar, $\Gamma_{\mathcal{C}}$ (or $\mathcal{C}$) vanishes or not should not affect $U_h$. The problem may be traced to be that the applied Gibbs distribution is not equally applicable to $\boldsymbol{u}_h$ and $u_z$ (a precise description of the physical measure of such a system is beyond the state-of-the-art of nonlinear science), or that the notion of statistical correlation does not reflect the dynamical relation of passiveness or activeness. Also, with $\Gamma_{\mathcal{C}} \neq 0$, there would be the opportunity of the concentration of $\mathcal{Z}$ at the smallest $k_h$ with equaipartition at large $k_h$ (c.f., Fig. 2), but, without $\Gamma_{\mathcal{C}}$, there is only the conventional energy equipartition for $\mathcal{Z}$. In such a situation, the appropriate coupling or feedback with the small $u_{z,z}$ perturbation helps the calculation making more sense. For instance, in some appropriate situation such partial coupling leads to the possibility of the transfer to large-scale $\mathcal{Z}$. Note that the helical absolute equilibria have recently been made to be of more physical illumination concerning chirality selection and amplification[14] without the above subtle passive-v.s.-active issue. Now, for convenience, we assume in the following remarks that just as the Rossby wave, the wave frequency $\omega \propto k_z$. It is then possible to show, by rewriting our Eq. (20) as $u_z = u_z(x, y, \zeta z, t, \zeta t, \zeta^2 t, ...)$ and with a multiple-scale expansion analysis, that, many-mode ('many'>3) resonances does be capable of feeding the 2D mode of $u_z$ with $k_z = 0$, similar to the zonal flow generation mechanism. Our point however is that before $t = O(1/\zeta^2)$ for the $k_z \neq 0$ modes of $u_z$ to feed through quartet resonance the $k_z = 0$ mode,[15] the back reaction already makes the helicity effective in the nonlinear relaxation: Indeed, the third term of Eq. (12), assumed to be critically balanced[16] with the second terms of both Eqs. (12 and 11), indicates that the relaxation time is of the order of the triad resonant time $O(1/\zeta)$ during which, in the context of zonal flow, "no major [triad] resonant rectification occurs."[8] Note that the relaxation towards the helical absolute equilibrium with $\mathcal{Z}$ concentrated at large scales should not induce inverse cascade/transfer just among the $k_z = 0$ modes, otherwise we would have the pathology as mentioned in the beginning of this paragraph. So, at such relaxation time scale, the only possible route is that $u_z$-waves deposit energy into large-scale $u_z$-vortex but then extract basically the same amount from other vortex modes (of $u_z$ at small scales, or, so far not excludable, of $\boldsymbol{u}_h$ at any scales). Our theory does not have any contradiction with current numerical results, for example, Figs. 6 and 7 of Ref. 17, although the information presented by the latter is still far from enough to make a claim of supporting all our predictions.

It is possible to extend the analyses to situations of higher dimensions or more (electro-magnetic) fields,[18] including those two-fluid models of neutral and plasma flows mentioned in the footnote 4. And, finally, we remark that our decomposition may be of value in rheology theory in the fashion of constructing constitutive equations, using the decompositions (of Giesekus-Harnoy-Drouot or Wedgewood[1], c.f., also other references there in the latter) beyond the corotational models:[19] It remains to first specify the rotation rate of the RSF coordinate frame and to find the objective part(s). However, the RSF of (1) rotating around the $z$ axis is still RSF, thus extra constraint is still wanted to settle down on a unique $Q$, which, among others, will be left for future studies and which could be of parallel value for turbulence modeling.[20]

[1] L. E. Wedgewood, Rheol Acta **38**, 91 (1999).
[2] I. Schur, Math Ann **66**, 488 (1909).
[3] H. K. Moffatt & A. Tsinober, Annu. Rev. Fluid Mech. 24, 281–312 (1992).
[4] In terms of differential geometry language with Lie derivative $L_{\boldsymbol{u}}$ with respect to $\boldsymbol{u}$, the 'frozen-in law' is translated to 'vorticity (either as a vector or a 2-form) being Lie-carried', $\partial_t \boldsymbol{\omega} + L_{\boldsymbol{u}} \boldsymbol{\omega} = 0$: c.f., T. Tao, [Ann. PDE **2**, 9 (2016)] for a recent formulation also in Euclidean space. Although we are here also focusing on incompressible case, the barotropic situation may be similarly treated. Using such a formulation is also because, like Tao's modified model, we have below a vorticity, say, $\boldsymbol{\omega}^{(2)}$ Lie-carried by another flow $\boldsymbol{u}$, instead of $\boldsymbol{u}^{(2)}$, and because we want to indicate that recent discovery of the so-called 'generalised Cauchy invariants equation' by N. Besse and U. Frisch [J. Fluid Mech. **825**, 412 (2017)], including the extension to non-ideal flows [J.-Z. Zhu, Phys. Fluids, **30** (3); see also, arXiv:1709.03356], carries over to current discussions, which however is not pursued here.
[5] C. J. Keylock and S. Tian et al. have applied RSF in their reseptive studies of flow structures (private communications in 2017 and 2018).
[6] The local RSF could be helpful for clarifying the *global regularity* issue. And, this is reminiscent of the locally inertial coordinates, supporting the *principle of equivalence* in general relativiy. We can also work with other vector gradients whose antisymmetric part correspond to the magnetic field, the electric current or else, and may formulate a kind of principle of equivalence for the specific dynamics.
[7] H. K. Moffatt, The degree of knottedness of tangled vortex lines. J. Fluid Mech. **35**, 117–129 (1969).
[8] H. P. Greenspan, The degree of knottedness of tangled vortex lines. J. Fluid Mech. **36**, 257–264 (1969).
[9] M. C. Marcetti et al., Hydrodynamics of soft active matter". Rev. Mod. Phys. **85**, 1143–1189 (2013).
[10] P. R. Cromwell, Knots and Links. Cambridge University Press (2004).
[11] T.-D. Lee, On some statistical properties of hydrodynamic and hydromagnetic fields. Q. Appl. Math. **10**, 69–74 (1952).
[12] R. H. Kraichnan, Inertial ranges in two-dimensional turbulence. Phys. Fluids **10**, 1417–1423 (1967).

[13] R. H. Kraichnan, Helical turbulence and absolute equilibrium. J. Fluid Mech. **59**, 745–752 (1973).

[14] J.-Z. Zhu, Chirality, extended magnetohydrodynamics statistics and topological constraints for solar wind turbulence. Monthly Notices of the Royal Astronomical Society **470**, L87 (2017); and references therein.

[15] A. C. Newell, J. Fluid Mech. **35**, 255–271 (1969).

[16] E.g., P. Goldreich and S. Sridhar, [Astrophys. J., **438**, 763 (1995)] whose idea has been extended to other fluid dynamics, including incompressible rotating and stratified turbulence [S. V. Nazarenko and A. A. Schekochihin, J. Fluid Mech. **677**, 134 (2011)] and compressible flows [J.-Z. Zhu, J. Fluid Mech. **787**, 440 (2016)].

[17] L. Bourouiba, D. N. Straub and M. L. Waite, J. Fluid Mech. **690**, 129–147 (2012).

[18] D. Montgomery & L. Turner, Two-and-a-half-dimensional magnetohydrodynamic turbulence. Phys. Fluids **25**, 345 (1982).

[19] R. B. Bird, O. Hassager and S. I. Abdel-Khalik, AIChE Journal, **20**, 1041 (1974).

[20] A D. Kirwan, Jr., Fluids, **1**, 3 (2016).



\section{Introduction}
\label{intro}
%\subsection{General background of two-dimensional-three-component dynamics and $\mathcal{C}$}
The governing equations for our two-dimensional (2D) passive scalar
turbulence problem are
\begin{eqnarray}
% \nonumber to remove numbering (before each equation)
  \partial_t \theta + \bm{v}\cdot\nabla\theta&=& \kappa\nabla^2\theta+f_{\theta},\label{eq:p1} \\
  \partial_t \bm{v} + \bm{v}\cdot \nabla \bm{v}&=& -\nabla P+\nu \nabla^2\bm{v}+f_{\bm{v}}, \ \nabla \cdot \bm{v}=0, \label{eq:p3}
\end{eqnarray}
where $f_{\bullet}$ denotes forcing. The scalar $\theta$ is \textit{passive}
because of no back-reaction onto $\bm{v}$ with $f_{\bm{v}}$ independent on
$\theta$ (otherwise \textit{active}), and $\bm{v}$ is the classical
incompressible 2D Navier-Stokes velocity field with $f_{\bm{v}}$ independent
of $\bm{v}$, in particular with $f_{\bm{v}}=0$ and that the pressure $P$
being related to $\bm{v}$ by the Poisson equation $-\partial_{\alpha\alpha}
P=\partial_{\beta\gamma}(v_{\beta}v_{\gamma})$. Curling (\ref{eq:p3}), we
have
\begin{equation}\label{eq:p2}
  \partial_t \zeta + \bm{v}\cdot\nabla\zeta = \nu\nabla^2\zeta+f_{\zeta},%: \ \bm{\zeta}=\nabla \times \bm{v}.
\end{equation}
%Note that, with $\bm{z}\perp \bm{v}$,
with the vertical vorticity $\zeta \bm{z}=\bm{\zeta}=\nabla \times \bm{v}$,
which is intrinsically different to the passive scalar in Eq. (\ref{eq:p1}).
The full 3D incompressible Navier-Stokes equations with $\partial_z=0$,
i.e., depending only on $x$ and $y$ coordinates, or alternatively, averaged over $z$, becomes such a 2D-three-component (2D3C) system and the vertical velocity $\bm{u}_z=u_z \bm{z}=\theta\bm{z}$ in the (unit) $\bm{z}$ direction is passively advected by the horizontal velocity $\bm{u}_h$ in the $x-y$ plane, thus a problem of 2D passive scalar with unit Prandtl number $\nu/\kappa$%(the ratio between viscosity and diffusivity)
. [Note that we have used $\bm{u}$ to represent the general velocity field,
instead of $\bm{v}$ for that in the $x-y$ plane; that is,
$\bm{u}_h=\bm{v}$.] The problem we focused on is the possibility of
different spectral transfer dynamics of the passive scalar energy like
various active scalars, with multidisciplinary physical relevance. Starting
from this section, we will subsequently present collectively four favorable
arguments. There are new ideas and/or techniques in each argument, which
requires careful reasoning and thus somehow (re)formulating the relevant
background materials with reasonable details. Experienced readers can jump
to the end of this section for a summary of our arguments and then, without
strictly following our formulation of the problems, selectively look for our
reasoning in each section.

2D3C (dominant) dynamics also emerge in the quasi-static magnetohydrodynamics
(QSMHD --- cf., e.g., Favier et al. \cite{FavierETCjfm11} and references therein)
due to extra linear Ohmic anisotropic dissipation operator, effectively different %constant-for-all-$\bm{k}$
damping rate $\propto cos^2\theta=k_z^2/k^2$ on different cone surfaces of
$\bm{k}$s forming different angles $\theta$s with the background magnetic field
direction $\bm{z}$ (```conical' Joule dissipation effect'' responsible for the `directional'
anisotropy \cite{Cambon90} and thus two-dimensionalization \cite{Moffatt67,Cambon90}).
%: The early linear (rapid distortion) theory \cite{Moffatt67} was advanced with a nonlinear closure \cite{Cambon90} adapted from that for rotating flows \cite{CambonJacquinJFM89}, and recently there appear other linear and nonlinear theories to account the dynamics as reviewed by Favier et al. \cite{FavierETCjfm11}. %Note that the study of Favier et al. \cite{FavierETCjfm11} ignores the 2D3C helicity effect which however is crucial to our theoretical discussions. %The QSMHD anisotropic damping rate on each $\theta$-cone surface is constant for all $\bm{k}$ there, just like the Eckman friction, though with distinct physical origin.
Yet another situation, in which 2D3C dynamics can arise and that also can be reformulated to be due to anisotropic dissipation,
is the so-called thin-layer flows, and there have been documentations on the transitions from
2D to 3D or from inverse- to co-existing/split and to forward cascades in the simulations
with periodic boundary conditions, not `bounded' or `without boundaries'
(see, e.g., Smith et al. \cite{SmithChasnovWaleffe96} and Celani et al. \cite{CMVprl10}).
Since, to our best knowledge, there is not any formulation for the two-dimensionalization of this
latter intuitive case, let us show here that simple scale-normalization argument makes the scenario quite transparent:
Suppose the box dimensions in $x$, $y$ and $z$ directions are $L_z R=L_x=L_y=L$, then by scale transformation $z'=zR$ %$L_{z'}=L_z/S=L$
it is direct to see, with corresponding re-scaling of the vertical velocity and forcing
(pressure can be canceled by incompressibility), that we have effectively new anisotropic
viscosity $\nu_{z'}R^{-2}=\nu_x=\nu_y=\nu$: The anisotropic viscosities lead to
anisotropic viscous scales, and the dynamics should be understood with corresponding
re-scaling of the forcing scales (if exist). The much larger $\nu_{z'}$ with large $R$
smoothes $z'$ dynamics more, leading to much smaller $\partial_{z'}$ (formally $\to 0$
with $R\to \infty$), i.e., asymptotically 2D3C. Especially, if the vertical forcing scale in
$z'$ is smaller than the vertical dissipation scale $\eta_{z'}$, then $z'$-direction motions
can hardly take effect. %Note however that, for $u_z \to u'_{z'}=u_z/R$
There are also other
mechanisms of two-dimensionalization, such as strong background magnetic field in a
plasma or strong rotation of a neutral fluid: Perturbative arguments for the
magnetohydrodynamics case for the former was given by Montgomery and Turner
\cite{MontgomeryTurnerPoF81}, say; the latter will be used and elaborated a bit more
later.
Note that quasi-two-dimensionalization in all these cases are quite obviously documented in experiments and numerical simulations and are easy to understand physically, though whether the pure 2D dynamics can be rigorously obtained in the corresponding limits may sometimes be subtle and even controvertible. %This article is about the physics of the 2D flow and its relevance to quasi-2D situations, so we don't need to get involved in the issue of whether the rigorous pure 2D limit can be available.

For the 2D3C dynamics, how the passive scalar is affected by the advecting
dynamics is measured by their cross-correlations. It turns out that the one
as the integral over the volume $\mathcal{V}$ of the $\bm{r}$ ($=[x,y]$)
space of the passive scalar and the vertical vorticity,
\begin{equation}\label{eq:crosscorrelation}
\mathcal{C}=2\langle\zeta\theta\rangle=\frac{2}{\mathcal{V}}\int \zeta\theta d\bm{r},
\end{equation}
is an ideal dynamical quadratic invariant (at least formally).
[With the ergodicity assumption, the volume average equals the statistical average, thus sharing the same notation $\langle \cdot \rangle$%, i.e., $\langle \cdot \rangle=\int \cdot d\bm{r}$
. The factor of $2$ is simply for later convenience.] Due to the fact that both $\theta$ and $\zeta$ are Lagrangian invariants, the multiplication of respectively any function of their own is also formally a Lagrangian invariant. In this sense, this 2D3C situation appears to be quite unique among all the $d$-dimensional problems. %, to iterate, stemming from the Lagrangian invariance of $\bm{\zeta}$.
This $\mathcal{C}$ is quadratic and `rugged' (see below). %is a rugged quadratic invariant and
It actually is the reduced helicity \cite{M69}; see also Sec. \ref{sec:AE} where %a speculation on the relevance to cyclogenesis is made but
the relation and difference to the `vertical/horizontal helicity (density)'
frequently referred to in weather science are noted.
%Literatures usually state that absolute-equilibrium passive scalar energy is equipartitioned, without any large-scale concentration, thus a forward turbulent cascade, which calls for a reexamination of this very basic point as a first step towards the full problem.
%So, we may simply identify the cross-correlation $\mathcal{C}$ between the passive scalar and the vorticity of such a 2D3C problem as the reduced form of helicity, when the boundary conditions are appropriate \cite{Moffatt69}.
Helicity is an ideal rugged invariant, i.e., conserved with or without spectral truncation \cite{k73}. So, it is intriguing whether $\mathcal{C}$ is controllable and what its dynamical effects are. %: indeed we will analyze with absolute equilibrium argument and be led to the conjecture that $\mathcal{C}$ could cause inverse transfers of the variance of the passive scalar. %Absolute equilibrium is nevertheless not turbulence and the more systematic analysis of a dissipative and forced system is desirable.

Note that the probability notion of `statistical correlation' does not distinguish the asymmetrical dynamical dependence, which may bring subtleties into the problem of passive scalar with explicit dynamically asymmetrical dependence: the scalar depends on the advecting field, but not the reverse. Simple statistical/probabilistic description by itself is inadequate/incomplete for dynamical systems: Here, due to dynamical dependence of $\theta$ on $\zeta$, it is possible to have covariance between them. \textit{The covariance statistically, but not dynamically, also means linear dependence of $\zeta$ on $\theta$; also, from the joint Gaussian distribution of the applied Gibbs ensemble for the absolute equilibrium (see later discussions), zero covariance would indicate statistical independence between them.} Further more, a way to control $\mathcal{C}$ is to have $f_{\theta}$ be appropriately correlated to the advecting dynamics, say, a reasonable $\langle f_{\theta}\zeta \rangle$ which by itself however could result from dynamical dependence of $\zeta$ on $f_{\theta}$, which is not the case for passive scalar problem. [Statistical description in principle can be infinitely complex, with, for instance, infinitely-many-dimensional (conditional) probability distribution functions (PDFs) or moments, so by `\textit{simple}' the limitations of finiteness of the PDFs or moments are referred to.] \textit{So, $\mathcal{C}$ itself and its control present some ambiguity with the passive and active scalars; or, in other words, controlling $\mathcal{C}$ for the passive scalar bridges the problem to the active scalar one. However, according to the dynamics, as long as there is no back-reaction onto the advecting field, the problem is still of passive nature.} %The probabilistic approach is inadequate or incomplete, but is still sometimes illuminating.

Ref. \cite{CCMVnjp04}'s review, of active versus passive scalars due to the
progresses \cite{FGVrmp01} of the Kraichnan model \cite{k68}, revisits the
inverse cascade of the active magnetic potential energy (the root mean
square of the `flux function', the vertical component of the vectorial
potential) of two-dimensional (2D) magnetohydrodynamics (MHD)
\cite{FyfeMontgomery76}. The relevant result may actually be used to make
the conjecture of the possibility of similar statistics for some special
passive scalars, beyond the very limited Kraichnan model and with some
similar necessary mechanisms of 2D MHD, which constitutes the important
point of this study. Our study will also be assisted with the analysis of
turbulence in a rotating frame of coordinates. We know that in the
small-Rossby-number limit, \textit{in particular situations and regimes},
the system contains a self-autonomous 2D3C subdynamics with the
vertically-averaged vertical velocity being a passive scalar. In general,
the pure 2D limit may not be unconditionally reached: The relevant time
scales for different non-resonant modes are continuously
\textit{non-uniform}, depending on how `near' to the resonant condition they
are \cite{Newell69}, which, together with the nonlinearity of near-resonant
interactions (unlike the linear an-isotropic damping in QSMHD), makes the
decoupling issue extremely subtle. For example, Chen et al. \cite{CCEH05}
showed that the long-time errors between the vertically-averaged results and
the pure 2D one grow, `implying non-resonant effects' and `is consistent
with' closure theories \cite{CambonRubinsteinGodeferd04}. And, Smith and Lee
\cite{SmithLee05} found, by examining the near-resonant, near-2D, pure-2D
and full-3D interactions, that near-resonant but not near-2D interactions
were responsible for the special 2D large-scale properties of 3D rotating
flows, that distinguished from the pure 2D ones, such as the steeper energy
spectra and the dominance of cyclones over anticyclones. They also noted
that the pure 2D interactions are necessary for the generation of 2D
large-scales. Of course, there may be subtle differences between the
geometries of a finite cyclic box and an infinite domain, and there may be
issues of discreteness and resolution effects, as noted by a series of
interesting works of Cambon and collaborators
\cite{CambonJacquinJFM89,CambonMansourGodeferdJFM97,CambonRubinsteinGodeferd04}
and as addressed, e.g., by Bourouiba and co-workers
\cite{BourouibaBartelloJFM07,BourouibaPRE08}. Recently there are numerical
simulations with \textit{intermediate}/\textit{moderate} Rossby numbers by
two independent working groups \cite{CCEH05,BourouibaJFM12}, showing indeed
large-scale formation of the vertically-averaged vertical velocity, while
Chen et al. \cite{CCEH05} noticed that simulations with even smaller Rossby
number do not show the corresponding inverse transfers. Bourouiba and
Bartello \cite{BourouibaBartelloJFM07} also particularly identified and
examined the \textit{`intermediate-Rossby-number' regime} which presents
special properties and to which previous relevant simulations by Smith and
collaborators and Chen et al. \cite{SW99,CCEH05,SmithLee05} were claimed to
belong. To try to understand the role of the 2D3C subdynamics embedded in
the full 3D3C system, in Sec. \ref{sec:RT} we thus focus on this issue and
speculate the triggering of spontaneous chirality (mirror symmetry breaking)
only with sufficient wave-vortex coupling in this regime. A further
speculation of 2D3C dominated helical cyclogenesis will be proposed in Sec.
\ref{sec:Remark}. A modification of the Kraichnan model with preliminary
calculations will also be presented in Sec. \ref{sec:ControlC} to bring
further insights for possible systematic solution of the problem. The
absolute equilibrium analysis of this work was originally a `byproduct' of
Ref. \cite{hydrochirality} and was meant to be added to it at its late
stage, particularly addressing the numerical reports of Refs.
\cite{map09,MP10}, so a relevant briefing, but with extra perspectives, is
given in the Appendix.

In summary, our main objective is to clarify the conjecture of the possible
non-universal genuine transfer directions of two-dimensional (2D) passive
scalar energy, given the inversely cascaded (to large scales) advecting
velocity, and collectively four arguments are used. In the following
sections, we will present \textit{1)}, the reasoning with the comparison
between passive and inversely-cascaded active scalars, given in this
introductory discussion, and we will offer \textit{2)}, the
absolute-equilibrium spectra shown concentration of 2D passive scalar energy
at both large- and small-scale ends, \textit{3)}, the detailed analysis and
explanation of the inverse transfers of the vertically averaged vertical
velocity in moderate-Ro rotating flows, together with \textit{4)},
preliminary calculation of a modified Kraichnan model showing how the
details of the passive scalar pumping mechanism matters. Remarks on helicity
in other situations and `vertical helicity' in hazardous weather also
follow.

\section{Reflection on active and passive fields: a new
perspective}
%Advances of passive scalar from the Kraichnan model \cite{k68} were obtained in the end of the last century and systematically reviewed at the very beginning of the millennium \cite{FGVrmp01}, but at this dawn some further progresses were made by comparing the passive and active ones with the same generic advecting field (see, e.g., Refs. \cite{CMMVprl02,CCMVprl02,CCGPepl02,CCGPpre03}) as summarized in
Ref. \cite{CCMVnjp04} reviews the studies on `active and passive fields face
to face', following the developments due to the progresses of the Kraichnan
model \cite{k68} (see Ref. \cite{FGVrmp01} and references therein). Not
surprisingly, active scalars such as 2D temperature/density, magnetic
potential and vorticity etc. are different - \textit{nonuniversal}, both in
the sense of the transfer directions and the inertial-range scaling laws. As
a sidenote, we remark that the conventional notion of \textit{universality}
in the Kraichnan model refers to the independence of the scaling exponents
on the `details of the pumping' of the passive scalar \cite{FGVrmp01}. The
pumping of the Kraichnan model is an independent Gaussian white-in-time
field with prescribed spacial correlation on which the scaling exponents are
independent. Beyond the Kraichnan model, the details of the pumping are
rich, including possible correlations with other variables, and may affect
the passive scalar statistics - the scaling exponents and others.
%, and it is remarked: ``...the passive scalar scaling laws are universal with respect to the injection mechanism...The basic property that makes passive scalar turbulence substantially simpler is the \textit{absence of} [emphasis added by us] statistical correlations between scalar forcing and carrier flow. On the contrary, the hallmark of active scalars is the functional dependence of velocity on the scalar field and, thus, on active scalar pumping....This poses the problem of universality in active scalar turbulence: if the forcing is capable of influencing the velocity dynamics, how can scaling exponents \textit{be universal with respect to the details of the injection mechanism} [emphasis added by us]?''
The interesting behaviors of active scalars, such as the inverse cascade of
the magnetic potential energy of 2D MHD, is related to the statistical
correlation of the pumping with the particle/tracer trajectories, i.e., the
advecting field. Such a correlation can be traced to the back-reaction
through $f_{\bm{v}}$ and in general is absent for the passive scalar, even
that advected by the same realization of velocity and pumped by
`statistically the same' (in the sense of its own probability distribution
function) but different independent realization of $f_{\theta}$
\cite{CCMVnjp04}. \textit{However, we note that even for the passive scalar
problem, given the same velocity field, there in principle can exist
realizations of the pumping, controlled artificially or by nature, to have
various statistical correlations with the particle trajectories, resulting
in, presumably, different statistics, including all the possible ones of
active and passive scalars in literatures.} Of course the passive scalar
inverse transfer, as the one of the active magnetic potential energy (the
root mean square of the `flux function', the vertical component of the
vectorial potential) of two-dimensional (2D) magnetohydrodynamics (MHD),
first predicted from the absolute equilibrium argument
\cite{FyfeMontgomery76}, is in the list. As noted in the end of the last
paragraph, this may require some adjustments of the traditional
understanding of passive scalar problem but is not really complete new: for
instance, Holzer and Siggia \cite{HolzerSiggia94} took the pumping be the
velocity projected onto a constant vector (taken to be the background
gradient of the scalar in Ref. \cite{HolzerSiggia94}). Thus the above
analysis already leads us to the conjecture of nonuniversal transfer
directions of passive scalars, especially the possibility of inverse
cascade/transfer of 2D passive scalar which will be further augmented with
absolute-equilibrium analysis by novel inclusion of the dynamical effects of
$\mathcal{C}$ (Sec. \ref{sec:AE}). As another sidenote, since we come to the
issue of inverse cascade/transfer of the passive scalar, we remark that its
existence or not is not necessarily related to the existence or absence of
the so-called dissipative anomaly, i.e., the persistent dissipation in the
vanishing diffusivity: With dissipative anomaly, it only means that some
energy is dissipated at small scales, which does not necessarily excludes
inverse cascade/transfer at large scales. [And, without dissipative anomaly,
energy could resides in some regime, which does not necessarily imply
inverse cascade/transfer.]

\section{$\mathcal{C}$-containing absolute equilibrium}\label{sec:AE}
For a 2D3C passive scalar advection the \textit{rugged} invariants are the
\textit{(horizontal) kinetic energy}
$\mathcal{E}=\langle u_h^2\rangle%=\int u_h^2dxdy/\mathcal{V}}
,$ the \textit{enstrophy}
$\mathcal{W}=\langle\zeta^2\rangle%=\int (\nabla\times\bm{u}_h)^2dxdy/\mathcal{V}
,$ and the \textit{passive-scalar (vertical) energy}
$\mathcal{Z}=\langle\theta^2\rangle=\langle u_z^2\rangle,$ besides
$\mathcal{C}$, Eq. (\ref{eq:crosscorrelation}). We now show that the
cross-correlation $\mathcal{C}$ is equivalent to the well-known invariant
helicity \cite{M69} \textit{under appropriate conditions}: With
$\bm{\zeta}=\nabla\times \bm{u}_h$ and $\theta=u_z$, we have
$\int \int \zeta\theta dxdy=\int \int \nabla\times \bm{u}_h\cdot \bm{u}_z dxdy.$ % is also conserved.
One can easily check by integration by parts, assuming vanishing velocity at
the boundary, say, infinity, or by assuming periodic boundary condition,
that
$$\int \int \nabla\times \bm{u}_h\cdot \bm{u}_z dxdy=-\int \int\bm{z}\times\nabla u_z \cdot \bm{u}_h dxdy.$$
Thus now $\mathcal{C}=2\langle \zeta\theta \rangle$ is nothing but the
reduced form of the well-known helicity, since, when $\partial_z\bm{u}=0$,
the integrand of helicity $$\nabla \times \bm{u} \cdot \bm{u}=\nabla\times
\bm{u}_h\cdot \bm{u}_z-\bm{z}\times\nabla u_z \cdot \bm{u}_h.$$ [And, then
issues such as the (detailed) conservation laws of $\mathcal{C}$ etc. are
reduced to what we know how to solve from \cite{k73}.] Note that \textit{the
local density of $\mathcal{C}$ is not $2\theta \zeta$ but $$\theta \zeta
-\bm{z}\times\nabla \theta \cdot \bm{u}_h.$$}

For $\bm{u}$ in a cyclic box with dimension $2\pi$, we have the Fourier
representation $\bm{u}(\bm{r})=\sum_{\bm{k}}
\hat{\bm{u}}_{\bm{k}}\exp\{\hat{i}\bm{k}\cdot\bm{r}\}$ and further expansion
with the self-evident notations $\bm{k}_h$ and $\bm{k}_z$ (following those
of the physical-space variables ---  all complex variables are consistently
wearing hats in this paper and $\hat{i}$ is the pure imaginary unit): For
the incompressible 2D flow, denoting the horizontal wavevector by
$\bm{k}_h$, we have $\hat{\bm{u}}_h\cdot\bm{k}_h=0$ and are left with only
one degree of freedom in the direction
$\bm{h}(\bm{k}_h)=\bm{z}\times\bm{k}_h/k_h$ for $\hat{\bm{u}}_h$; so,
writing $\hat{\bm{u}}_h=\hat{u}_h(\bm{k}_h)\bm{h}(\bm{k}_h)$, we have
$$\mathcal{C}=\tilde{\sum}\hat{i}k_h(\hat{u}_h\hat{u}_z^*-c.c.), \ \mathcal{E}=\tilde{\sum}|\hat{u}_h|^2, \ \mathcal{W}=\tilde{\sum}k_h^2|\hat{u}_h|^2 \ \textnormal{and} \ \mathcal{Z}=\tilde{\sum}|\hat{u}_z|^2,$$
where we have applied the conventional Galerkin truncation keeping only
modes with $k_{min} \le |\bm{k}_h| \le k_{max}; \ \textnormal{thus} \
\tilde{\sum}\triangleq \sum_{k_{min} \le |\bm{k}_h| \le k_{max}}.$
`$\triangleq$' means definition, and for simplicity, we have omitted the
conventional factors of $1/2$ which can be absorbed into the (inverse)
temperature parameters (see below). [Alternatively, following the notion of
helical inertial waves of rotating flows \cite{c61}, the helical
representation (see below) in 2D reduces to $\hat{v}_{+}=-\hat{v}_{-}$; so,
one may use $\hat{\bm{u}}_h=\hat{u}'_h(\bm{k}_h) \hat{\bm{h}}'(\bm{k}_h)$
with $\hat{\bm{h}}'(\bm{k}_h)=\hat{i}\bm{z}\times\bm{k}_h/k_h$, and yet
another way for interested readers to exercise and check the calculations is
starting with the Fourier expansion of the familiar stream function
$\bm{\psi}=\psi \bm{z}$ for the horizontal flow, with $\bm{v}=-\nabla \times
\bm{\psi}$ and $\bm{\zeta}=\nabla^2\bm{\psi}$.] It is direct to check that
all these quadratic invariants are conserved in detail for each interacting
triad and that, together with their global conservation laws, the quadratic
and diagonal properties ensure their \textit{ruggedness} after arbitrary
truncations \cite{k73}, which justifies respecting all of them in the
statistical treatment.

For the Galerkin-truncated inviscid and diffusionless system, triadic
interactions of the modes through the convolution in the convective terms in
general will present chaotic dynamics and lead to thermalization, i.e.,
approaching the thermal equilibrium. Indeed, as shown by Lee \cite{Lee1952}
for 3D incompressible hydrodynamics, the ordinary dynamical system in terms
of the Fourier modes satisfies the Liouville theorem which ensures an
invariant measure. Similar result was also obtained by Hopf \cite{Hopf52}
who applied functional calculus to formally derive it, but without
explicitly introducing the Galerkin truncation. [The classical wisdom is
that the physical measure may be accurately represented by the Gibbs state,
at least for low order moments.] And, such fundamental dynamics should play
a central role for many properties of the dissipative (and diffusive)
turbulence; the spectral transfer property should be signatured by such an
internal drive of thermalization. Recently Moffatt \cite{Moffatt14} studied
the single-triad interactions, but for statistical considerations we in
general need ``many'' \cite{Orszag} triads (still, the
\textit{thermalization assumption} fails for however many triads if the
superposed Fourier modes result in vanishing nonlinearity
\cite{sphmRapids}); otherwise, other unknown invariants might emerge or
ergodicity might seriously break down \cite{Orszag,JonLee79}. Besides other
possible footprints, the statistical absolute equilibria set up the aims
towards which the distribution of the spectra tend to relax \cite{k67,k73},
thus some clues of cascade directions may be obtained. However, before
proceeding, we should point out that, although successful in many aspects,
such an approach nevertheless misses many other ingredients of the dynamics,
which makes it in general difficult to conclude from the results very
accurately and firmly about turbulence; and, a caveat particular to the
passive scalar problem is that the calculations can not exploit the
passiveness of the scalar, as is a particular case of the
inadequacy/incompleteness of simple statistical description remarked in the
introductory discussion. Some finer statistical treatment than the
conventional microcanonical ensemble may be necessary to be more precise. We
should view the absolute equilibrium calculation as a unique mathematical
treatment to expose fundamental aspects of the dynamics. Recently, the idea
has been extended to study explicitly realizable fractally decimated
\cite{FrischETCprl12} and also some chirally selected (\cite{ZhuPoF14} and
references therein) Navier-Stokes systems, and, besides transfer directions,
insights about the isotropic polarization issue and similarly
multiple-constraint nonequilibrium dynamical ensembles
\cite{SheJackson,nonir} can be motivated \cite{ZhuJFM15}.

Introducing corresponding Lagrange multipliers or the (inverse) temperature parameters $\Gamma_{\bullet}$% to form the constant of motion%$\mathcal{S}=\gamma_{\mathcal{C}}\mathcal{C}+\gamma_{\mathcal{E}}\mathcal{E}+\gamma_{\mathcal{W}}\mathcal{W}+\gamma_{\mathcal{Z}}\mathcal{Z}$
, we now apply the Gibbs distribution, i.e., the canonical ensemble with
constant temperature parameters assigning a probability $\mathcal{P}$ to
each microstate
$$\mathcal{P} \sim
\exp\{-(\Gamma_{\mathcal{C}}\mathcal{C}+\Gamma_{\mathcal{E}}\mathcal{E}+\Gamma_{\mathcal{W}}\mathcal{W}+\Gamma_{\mathcal{Z}}\mathcal{Z})\},$$
to obtain the modal spectral densities, $U_h$ of $\mathcal{E}$, $W$ of
$\mathcal{W}$, $Q_{\mathcal{C}}$ of $\mathcal{C}$ and $U_z$ of
$\mathcal{Z}$:
%\begin{widetext}
\begin{eqnarray}
\!\!\!\!\!\!\!\!\!\!\!\!\!\!\!\!  \text{with}\ \ D=\Gamma_{\mathcal{E}}\Gamma_{\mathcal{Z}}+(\Gamma_{\mathcal{W}}\Gamma_{\mathcal{Z}}-\Gamma_{\mathcal{C}}^2)k_h^2>0, \   U_h\triangleq \langle |\hat{u}_h|^2 \rangle= \frac{\Gamma_{\mathcal{Z}}}{D{\color{red} %|_{\Gamma_{\mathcal{C}}=0}
}}, \ W=k_h^2 U_h, \label{eq:2D3CU}  \\
\!\!\!\!\!\!\!\!\!\!\!\!\!\!\!\!  Q_{\mathcal{C}}\triangleq \langle \hat{i}k_h\hat{u}_h\hat{u}_z^* \rangle+c.c.=\langle \hat{\zeta}\hat{\theta}^*\rangle+c.c.= \frac{-2\Gamma_{\mathcal{C}}k_h^2}{D}, \ U_z\triangleq %\langle |\hat{u}_z|^2 \rangle=
\langle |\hat{\theta}|^2 \rangle= %\frac{\Gamma_{\mathcal{E}}+\Gamma_{\mathcal{W}}k_h^2}{D}
\frac{1}{\Gamma_{\mathcal{Z}}}+\frac{\Gamma^2_{\mathcal{C}}k^2_h}{\Gamma_{\mathcal{Z}}D}=\frac{\Gamma_{\mathcal{E}}+\Gamma_{\mathcal{W}}k_h^2}{D}. \label{eq:2D3CQ}
\end{eqnarray}
%\end{widetext}
Note that $D>0$ and $\Gamma_{\mathcal{Z}}>0$ from the realizability, and
there are two particularly interesting situations:
\begin{itemize}
  \item One is that when $\mathcal{C}=0=\Gamma_\mathcal{C}$% (or alternatively $\Gamma_{\mathcal{Z}} \to \infty$)
  , we recover the absolute equilibrium spectra commonly in people's minds; in particular, $U_z$ is equipartitioned and $U_h$ is the Kraichnan 2D Euler absolute equilibrium energy (modal) spectral density. As is well-known (e.g., \cite{HolzerSiggia94,FGVrmp01}), equipartitioned $\mathcal{Z}$ indicates a forward cascade in the turbulent state of the passive scalar variance with diffusivity acting at the small scales \cite{CelaniETC00}. And, in the numerical simulations without helicity in the rotating frame at small Rossby numbers, it has been found \cite{BourouibaPoF08} that such equilibrium state is identifiable during the long-time (or metastable) transient state to the final isotropic state \cite{Lee1952,YKR02}.  %And, this can also correspond to the $Ro\to 0$ limit with completely decoupled pure 2D3Csubdynamics.
%  \item When $\Gamma_{\mathcal{W}}\Gamma_{\mathcal{Z}}-\Gamma_{\mathcal{C}}^2 \le 0$, the situation is similar to the case with $\Gamma_{\mathcal{C}}=0$ and $\Gamma_{\mathcal{W}}<0$, except that both $U_z$ and $U_{\mathcal{C}}$ now may concentrate at the largest $k_h$. This is an extension of a situation of 2D Euler by \cite{k67} with his $\alpha>0$ and $\beta<0$.
  \item The other is that of $\Gamma_{\mathcal{E}}<0$ and that an
      $\mathcal{E}$ condensation state. When
      $\Gamma_{\mathcal{W}}\Gamma_{\mathcal{Z}}-\Gamma_{\mathcal{C}}^2>0$,
      non-vanishing $\Gamma_{\mathcal{C}}$ does not qualitatively change
      the standard feature of 2D energy-enstrophy absolute equilibrium
      spectra and that the indication for inverse horizontal energy
      transfer still stands. So, for the 2D3C horizontal velocity
      statistics compared to the pure 2D one \cite{k67}, it is just a
      transformation of the temperature parameters, without changing the
      essential physical characters: For instance, the energy
      equipartition does not correspond to $\Gamma_{W}=0$ anymore, but
      instead to
      $\Gamma_{\mathcal{W}}\Gamma_{\mathcal{Z}}-\Gamma_{\mathcal{C}}^2=0$.
      Though $\mathcal{C}$ by itself does not distinguish passive or
      active correlation of $\theta$ with $\zeta$ (as remarked in the
      introductory discussion) and though $\Gamma_{\mathcal{C}}$ appears
      in $U_h$, here no effective artificial back-reaction of the passive
      scalar through $\mathcal{C}$ has taken place. That is, the applied
      Gibbs measure may still be a `physical' one (in the sense of
      numerical realizability) \cite{footnotePhysicalMeasure}.
      \textit{Now, $U_z$ for $\mathcal{Z}$ is not equipartitioned any
      more, and when $\mathcal{E}$ condensation is strong enough, a
      significant amount of $\mathcal{Z}$ also resides at large scales
      (c.f., Fig. \ref{fig:UhUz}), which, without other effective
      constraints for transfers, indicates the possibility of transferring
      $\mathcal{Z}$ to form large-scale $\mathcal{Z}$-structures.}
        \begin{figure}
          % Requires \usepackage{graphicx}
          \begin{center}
          \includegraphics[width=0.45\textwidth]{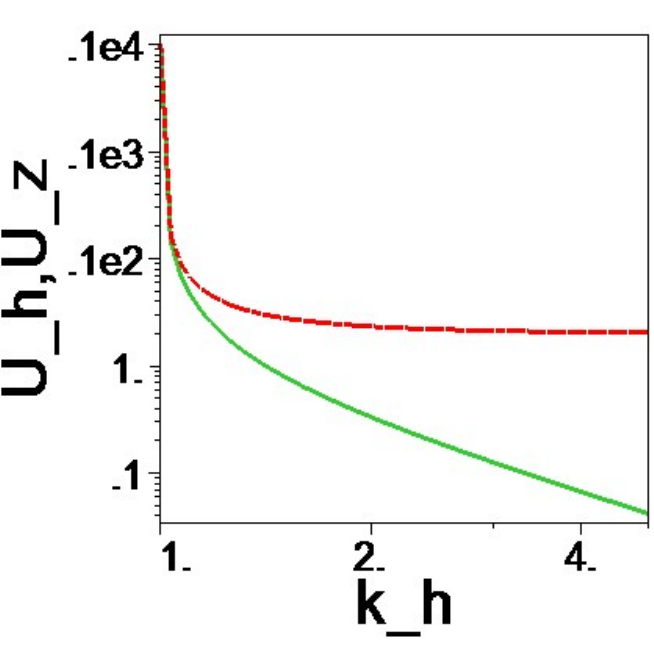}\\
          \caption{An example of the modal spectra concentrating at large scales ($k_{min}=1$): Solid line (blue) for $U_h$ and dash line (red) for $U_z$ are plotted with $\Gamma_{\mathcal{E}}=-1$, $\Gamma_{\mathcal{Z}}=1$, $\Gamma_{\mathcal{W}}=2.001$ and $\Gamma_{\mathcal{C}}=-1$.}\label{fig:UhUz}
          \end{center}
        \end{figure}
      Such an indication seems to be supported by the numerical measurements shown in Fig. 6 of \cite{CCEH05} and Figs. 6 and 7 of \cite{BourouibaJFM12}, which however is subtle since such data are not for the small-$Ro$ asymptotic 2D3C subdynamics but for intermediate Rossby number\textbf, requiring more discussions as given in the next section. Note that now whatever $\Gamma_{\mathcal{C}}$ or $\mathcal{C}$, large-$k_h$ $\mathcal{Z}$ spectrum is asymptotically an equipartition one, so the commonly accepted indication for forward cascade always exists. The (conjectured) inverse and forward cascades of passive scalar have nothing in conflict, because they belong to opposite scale regimes, at whose two respective ends the absolute-equilibrium spectra properties are used for the prediction. %Nevertheless, the strong dissipative of nature of $\mathcal{Z}$ and $\mathcal{C}$ at small scales, like $\mathcal{W}$, makes their inverse transfers at large scales highly non-trivial. %In plain words, the inverse transfer of the passive scalar energy at large scales under appropriate situations is not necessarily relevant to the issue of dissipative anomaly.
\end{itemize}

We remark that both $Q_{\mathcal{C}}$ and $U_z$ scale with large $k_h$
similar to $W$. The fact that the (weaker than $U_h$) condensation of $W$ at
small $k_h$ does not lead to inverse cascade of enstrophy is mainly due to
the mutual constraint $W=k_h^2U_h$ and the independent conservation laws of
energy and enstrophy which prohibit $\mathcal{W}$ from going together with
$\mathcal{E}$ totally to small $k_h$ \cite{k67}.
%But, compared to $U_h$, the weaker condensation at small $k_h$ and stronger emphasis at large $k_h$ (asymptotically equipartitioned) of $U_{z}$ do require us exert some carefulness.\cite{FluxFootnote} Although we believe the condensation and the lack of other effective constraints may still cause inverse transfer to large scales, we are aware that the situation is more subtle than that of $\mathcal{E}$.
Actually, the emphasis at both large and small $k_h$ of $U_{\mathcal{Z}}$
leads us to the conjecture that $U_h$ could genuinely cascade in a split way
to both  large and small scales: When the passive scalar energy is pumped at
some intermediate scale(s) at some appropriate level so that small-scale
dissipation removes part of the source cascading forwardly, a large-scale
``friction'' might act to balance the left energy transferred inversely.
That is, the small-scale asymptotic is similar to the enstrophy dynamics,
while the large-scale transfer can be similar to the kinetic energy of the
2D velocity: \textit{energy-enstrophy duality} of the 2D passive scalar with
effective $\mathcal{C}$. $Q_{\mathcal{C}}$ is neither effectively
constrained, thus $\mathcal{C}$ itself may have spectral transfer properties
similar to $\mathcal{Z}$, including the inversely-to-large-scale one. Of
course, the lack of effective constraint on them can also lead to their more
efficient dissipation at small scales, but we remind and iterate our remark
given in the introductory discussion that the existence or not of
(small-scale) dissipative anomaly has no necessary relevance to the
existence or absence of the genuine inverse cascade/transfer at large
scales. Absolute equilibrium is nevertheless different from the
nonequilibrium and dissipative turbulence, and all such possibilities from
purely the conservative triadic interactions, though genuine, may require
special realistic physical settings to persist, which deserves further
discussions.

We wind up this section by another sidenote that so far all kinds of
estimates of the fluxes need \textit{a priori} assumptions (say, on the
H\"older exponent or the ansatzes of the energy spectra themselves) of the
fields \cite{EyinkEuler08}; see in particular Sulem and
Frisch\cite{SulemFrischJFM75} for the passive scalar problem. So, the
inference of the turbulence fluxes from the zero-flux equilibrium is in a
sense more self-consistent, though less direct and quantitative.

\section{The inverse transfers of vertically-averaged vertical velocity variance/energy of rotating flows}\label{sec:RT}
\subsection{Some background of resonant wave theory and 2D3C sub-dynamics}
The fluid dynamics in the rotating frame of reference with uniform
$\bm{\Omega}=\Omega \bm{z}$ are governed by
\begin{equation}\label{eq:rtr}
    \frac{\partial\bm{u}}{\partial_t}+2\frac{\bm{z}\times\bm{u}}{Ro}=\bm{u}\times (\nabla\times\bm{u})-\nabla P + \frac{\nabla^2\bm{u}}{Re}; \ \nabla \cdot \bm{u}=0,
\end{equation}
where $Re$ is the Reynolds number and $Ro$ is the Rossby number
characterizing the (inverse) Coriolis force, and, where the centrifugal
force has been absorbed into the pressure $P$. This equation in the linear
inviscid limit admits inertial waves of the form\cite{c61}
%\begin{widetext}
%\begin{equation}\label{eq:iw}
$%\bm{u}=
\hat{\bm{h}}_{c_{\bm{k}}}
\exp\{\hat{i}(\bm{r}\cdot\bm{k}+\omega_{c_{\bm{k}}} t)\}$
%\end{equation}
%\end{widetext}
$\textnormal{with} \ \omega_{c_{\bm{k}}}=c_{\bm{k}}2\bm{k}\cdot\bm{z}/(kRo),
\
\hat{i}\bm{k}\times\hat{\bm{h}}_{c_{\bm{k}}}=c_{\bm{k}}k\hat{\bm{h}}_{c_{\bm{k}}}
\ \textnormal{and} \ c_{\bm{k}}^2=1=-\hat{i}^2.$ Here $c_{\bm{k}}=\pm$
designates chirality of the helical wave, i.e., the spiral directions along
$\bm{k}$. Such $\hat{\bm{h}}_{c_{\bm{k}}}$ are actually the eigenmode of the
curl operator and form the orthonormal
[$\hat{\bm{h}}_{c_{-\bm{k}}}=\hat{\bm{h}}_{c_{\bm{k}}}^*=\hat{\bm{h}}_{-c_{\bm{k}}}$
and
$\hat{\bm{h}}_{c1_{\bm{k}}}\cdot\hat{\bm{h}}_{c2_{\bm{k}}}=\delta_{{c1_{\bm{k}}},{c2_{\bm{k}}}}$]
bases for a transverse vector field $\bm{u}(\bm{r})$ in a cyclic box
\cite{CambonJacquinJFM89,W92}
%\begin{equation}\label{eq:FourierHelical}
$\bm{u}=\sum_{\bm{k}} \hat{\bm{u}}_{\bm{k}}e^{\hat{i}\bm{k}\cdot
\bm{r}}=\sum_{\bm{k},{c_{\bm{k}}}}
\hat{u}_{c_{\bm{k}}}\hat{\bm{h}}_{c_{\bm{k}}}e^{\hat{i}\bm{k}\cdot \bm{r}}$.
%\end{equation}
With such a representation, Eq. (\ref{eq:rtr}) transforms into \cite{W93}
\begin{eqnarray}
    (\partial_t -\hat{i}\omega_{c_{\bm{k}}}+\frac{k^2}{Re})\hat{u}_{c_{\bm{k}}}=
    \sum_{\bm{k}+\bm{p}+\bm{q}=0}\sum_{c_{\bm{p}},c_{\bm{q}}}\hat{C}_{c_{\bm{p}}c_{\bm{q}}}^{c_{\bm{k}}} \hat{u}^*_{c_{\bm{p}}}\hat{u}^*_{c_{\bm{q}}} \label{eq:uk}
        %=\frac{1}{4\hat{\bm{h}}_{c_{\bm{k}}}^*\times \hat{\bm{h}}_{c_{\bm{p}}}^* \cdot \hat{\bm{h}}_{c_{\bm{q}}}^*}
\end{eqnarray}
with $\hat{C}_{c_{\bm{p}}c_{\bm{q}}}^{c_{\bm{k}}}=\hat{\bm{h}}_{c_{\bm{k}}}^*\times \hat{\bm{h}}_{c_{\bm{p}}}^*\cdot \hat{\bm{h}}_{c_{\bm{q}}}^*(c_{\bm{q}}q-c_{\bm{p}}p)/2$ %. $\frac{\hat{C}_{c_{\bm{p}}c_{\bm{q}}}^{c_{\bm{k}}}}{c_{\bm{q}}q-c_{\bm{p}}p}=\frac{\hat{C}_{c_{\bm{q}}c_{\bm{k}}}^{c_{\bm{p}}}}{c_{\bm{k}}k-c_{\bm{q}}q}=\frac{\hat{C}_{c_{\bm{k}}c_{\bm{p}}}^{c_{\bm{q}}}}{c_{\bm{p}}p-c_{\bm{k}}k}$, resulting
whose symmetry properties (see later discussions) result in the detailed
conservation laws of energy and helicity among each triad
$\{[\pm\bm{k},c_{\bm{k}}];[\pm\bm{p},c_{\bm{p}}];[\pm\bm{q},c_{\bm{q}}]\}$
with $\bm{k}+\bm{p}+\bm{q}=0$.
%The \textit{mean} energy $\mathcal{E}=\sum_{c,\bm{k}}U^c(\bm{k})$ and helicity $\mathcal{C}=\sum_{\bm{k},c}Q^c(\bm{k})$ come from
%\begin{equation}\label{eq:QU}
%U^c(\bm{k})=cQ^c(\bm{k})/k=\langle|\hat{u}_c(\bm{k})|^2\rangle/2,
%\end{equation}
%where $\langle \bullet \rangle$ denotes the mean, per unit volume or by assumption of ergodicity, equally the statistical average.
When the Coriolis term goes to zero with $Ro\to\infty$, the system formally
reduces to that in the normal inertial frame; but, if
$\omega_{c_{\bm{k}}}=0$, {\it i.e.}, $k_z=\bm{k}\cdot\bm{z}=0$, the system
is seen to formally reduce to be 2D3C. In resonant wave theory, this 2D3C
Navier-Stokes is called the \textit{vortex/slow} modes, while the rest with
$k_z \neq 0$ is called 3D \textit{wave/fast} modes. A key issue in the
theory is about the \textit{decoupling}, \textit{recoupling} and their
\textit{interactions} of these parts. In the rapid rotation case, $Ro$ is
small and the inertial waves oscillate fast, which results in a
multiple-time-scale problem. Resonant interaction theory then assumes two
time scales ($t$ for slow and $\tau=t/Ro$ for fast dynamics) and the {\it
ansatz}
$\hat{u}_{c_{\bm{k}}}=\hat{V}_{c_{\bm{k}}}(t)\exp\{\hat{i}\omega_{c_{\bm{k}}}\tau\}$
which when brought into Eq. (\ref{eq:uk}) leads with asymptotic expansion to
the approximate averaged equation
\begin{equation}
    (\partial_t +\frac{k^2}{Re})\hat{V}_{c_{\bm{k}}}=
    \sum^{\omega_{c_{\bm{k}}}+\omega_{c_{\bm{p}}}+\omega_{c_{\bm{q}}}=0}_{\bm{k}+\bm{p}+\bm{q}=0}
    \sum_{c_{\bm{p}},c_{\bm{q}}}\hat{C}_{c_{\bm{p}}c_{\bm{q}}}^{c_{\bm{k}}} \hat{V}^*_{c_{\bm{p}}}\hat{V}^*_{c_{\bm{q}}}. \label{eq:secular}
\end{equation}
This equation is nothing but the Navier-Stokes in the inertial frame with
nonlinear interactions limited to the resonant modes. Two-dimensionalization
then may be argued at appropriate time regimes
\cite{W93,CambonRubinsteinGodeferd04,CCEH05}, as indeed confirmed by many
simulations. However, precise and detailed knowledge, especially in the
large Reynolds number limit, is far from clear. Note in particular that the
above resonant condition excludes the transfer of energy from (to) two fast
modes to (from) a slow mode; in other words, the 2D modes evolve
autonomously, though interact with fast modes as a catalyst. So, the 2D3C
modes evolve just as if the other 3D fast modes were truncated. Formally the
above results are true for $Ro \to 0$, and indeed it has been proved to be
valid for finite time with given finite $Re$ (see, Ref. \cite{CCEH05} and
references therein): There is the problem of exchanging limits among $t\to
\infty$, $Ro\to 0$ and $Re\to \infty$ now,\cite{footnote}
%\footnote{It may make it clearer to quote the remarks by \cite{CCEH05} that ``the present theorems effectively say nothing about the validity of the resonant wave theory at hight $Re$, for realistic values of the Rossby number'' and ``the asymptotic analysis does not determine the largest possible time'' for which theory is proved to be valid.}
and indeed numerical analyses do appear to confirm the validity in a finite
time with sufficiently small $Ro$ and large $Re$, while also indicating
breakdown beyond some intrinsic time interval. It is thus intriguing how the
theorem would be practically effective and useful for finite $Re$ and finite
$Ro$. And, what are the transfer dynamics relevant to the observed
interesting phenomenon, such as two-dimensionalization? Note that there may
be other subtleties for infinite domain case
\cite{CambonJacquinJFM89,CambonRubinsteinGodeferd04}, while discreteness and
resolution effects in simulations for flows in a cyclic rotating box have
been checked by Bourouiba \cite{BourouibaPRE08}.

\subsection{Theoretical observation and conjecture of spontaneous mirror symmetry breaking/chirality}
As a trivial interpolation, and as is well documented by numerical results
\cite{CCEH05,BourouibaPoF08,BourouibaJFM12,SmithLee05}, loosely speaking the
intermediate-$Ro$ dynamics is intermediate in between the large-$Ro$ and
small-$Ro$ ones, in the sense that resonant wave theory is partially working
with the slow modes incompletely decoupled from the fast ones; but, it has
also been found \cite{CCEH05,BourouibaBartelloJFM07,BourouibaJFM12} that
there are non-trivial non-monotonic properties which identify an
`intermediate-Ro regime', which is also our focus here. Bourouiba's
\cite{BourouibaPoF08} simulations, without considering $\mathcal{C}$, did
show that during the long-lived transient stage to the final fully 3D state
\cite{YKR02}, the 2D3C dynamics is clearly identifiable. Now there are other
important resonant and non-resonant coupling mechanisms which can kick in to
effectively make the helicity be a somewhat more active constraint, due to
finite $Ro$ and that the lack of complete decoupling, in which case the fast
modes may act as a bath to slowly modify the temperatures of the 2D3C
``thermometer'', which would make the $\mathcal{C}$-containing Gibbs
absolute equilibrium obtained in the last section more meaningful, in the
sense that the vertically-averaged vertical velocity can have slow feedback
onto the horizontal slow modes.
In \cite{CCEH05}, the forcing is acting at some small scales with a scheme depending on the velocity field itself, to give a constant energy injection in all three components and all fast or slow modes. Such a scheme does not impose or tell whether there is any helicity injected, neither for the full 3D Naiver-Stokes nor for the vertically-averaged dynamics (with the force also vertically averaged). Their Fig. 6b, with small Rossby number and that strong decoupling, says that there is probably not; their Fig. 6a for intermediate Rossby number indicates that the 2D3C subsystem probably obtain some helicity, either from the forcing or from the coupling with the fast modes; further examination of such data, especially those relevant to helicity (transfer), would be illuminating. Ref. \cite{BourouibaJFM12} has presented consistent results with intermediate Rossby number in their Figs. 6(c,d) and 7(c,d), however their forcing scheme like that of \cite{CCEH05}, but acting only on the inertial-wave modes, still tells nothing about the injection of cross-correlation or not. %The horizontal velocity of \cite{CelaniETC00} was specified to be in the inverse transfer range, but the forcing for the passive scalar was at large scales (again without any information about the injection of cross-correlation derivable from the schemes), thus not a setup for the inverse transfer of $\mathcal{Z}$.

%It could be possible that n
Neither Chen et al. \cite{CCEH05} nor Bourouiba et al. \cite{BourouibaJFM12}
explicitly injected helicity into the system by their forcing schemes,
which, to support our argument, might require \textit{spontaneous mirror
symmetry breaking/chirality} in the vortex and wave subsystems who exchange
helicities of opposite chiralities through the partial coupling: This in
principle can be further checked in the data. Saying ``spontaneous'' is
because the corresponding helicity equation from either chiral sector of Eq.
(\ref{eq:uk}) is not affected by the Coriolis term for either vortex or wave
subsystem, thus no explicit chiral symmetry breaking mechanism in the
dynamics. Also possible is that, after the initial spontaneous symmetry
breaking, the velocities at the forcing scale $k_f$ may adjust themselves to
make the forcing be helical. These are our further speculations that may
support their finding of the inverse transfer of $\mathcal{Z}$.

%The absolute equilibrium indication can also provide the basis for the discussions in Sec. 6 of \cite{BourouibaJFM12}:
The tendency towards the large-scale condensation spectra of $\mathcal{E}$ and $\mathcal{Z}$ can obtain energies directly through the wave-vortex coupling from the wave modes, besides the conventional inverse transfers to ``even larger'' scales within the vortex modes themselves; and, depending on the details (strength, ``location'' in scale space etc.) of the coupling which may be affected by the forcing schemes, inverse-energy and forward-enstrophy cascades, or some mixtures, are all possible. Indeed, Ref. \cite{BourouibaJFM12} found in their simulations that the `$33\to 2$' (wave-wave to vortex: they denote wave modes with `3' and horizontal vortex/slow modes with `2') transfer through the coupling was mostly at large scales, and, since \textit{their external forcing at small scales is only on the wave modes}, they proposed a forward enstrophy cascade of the vortex modes to explain the scaling exponent of $\mathcal{E}$ spectrum close to $-3$; they also found that `$33\to w$' (wave-wave to $w$, the vertically-averaged/slow-part of $u_z$) coupling feeded $\mathcal{Z}$ at medium to large scales while `$2w\to w$' transferred $\mathcal{Z}$ to the dissipation regime. This does not contradict the implication from our absolute equilibrium. It may be simply that some kind of ``inverse transfer'', even through the external channel of the wave-vortex coupling, of energies to large scales should be facilitated to support a large-scale energy condensation state. %; we will come back to this in the next subsection.
Interestingly, according to \cite{CenciniMGV11}, even if the small-scale
slow modes are also forced by the external forcing or by the slow-fast
coupling, a nonlinear superposition of forward-enstrophy and inverse-energy
cascades of slow modes also appears to support the observed atmospheric
$k^{-3}$ spectra among many other proposals (c.f., Ref. \cite{SW99}): Ref.
\cite{BourouibaJFM12} showed that the large-scale pumping can be provided by
the partial vortex-wave coupling, and the condensation absolute equilibrium
spectra may offer an intrinsic mechanism for such coupling to feed the
large-scale vortex modes; for more critical readers see below for more.

\subsection{Relevant questions on our conjecture and possible solutions}
One may fairly question: Why the even smaller-Ro (=0.0021) case of Chen et
al. \cite{CCEH05} does not show inverse transfer of $\mathcal{Z}$? And, the
$2w-w$ interactions are those responsible for the conservation laws used in
the absolute equilibrium calculation; then, why it was observed in
\cite{BourouibaJFM12} that the inverse transfer of $\mathcal{Z}$ was
provided by $33-w$ while the forward transfer was given there by $2w-w$?
`Devils' are in the dynamical details, which are beyond the general
conservation-law argument, and may provide the explanations: When the Rossby
number is too small, the coupling would be too weak to be able to trigger
the spontaneous mirror symmetry breaking (chirality) through the coupling
(dissipation might work jointly as the trigger --- note that the
equipartitions presented for the small-Ro Galerkin-truncated inviscid
simulations of Bourouiba \cite{BourouibaPoF08} indicate that there is no
spontaneous mirror symmetry breaking between that reasonably decoupled
subsystems.) And, of course, when the Rossby number is too large, the
coupling is too strong to identify the 2D3C subdynamics separately.
%Our result indicates the possibility, under the condition of sustained cross-correlation, that in the small-Rossby-number limit all the slow-mode energy goes to large scales, while that of fast modes all to small scales (the latter is supported by the absolute equilibrium of the fast modes, formally with no difference to that of \cite{k73} for the 3D Euler in the inertial frame.)
%Such discussion of turbulence based on the absolute equilibrium may sound quite of speculative nature, but given the consistency, besides the check of helicities in the subsystems proposed above, further Lagrangian inspections similar to those performed in Ref. \cite{EyinkNature13} should be suggested.
%Such a possibility, just as that for pure 2D passive scalar, requires a clever way to maintain $\mathcal{C}$ without being easily damped.
For the latter question, though it is true that Bourouiba et al. \cite{BourouibaJFM12} did not directly show $2w-w$ inverse transfer, one possible consistent dynamical scenario goes as follows. First of all, we need to make it clear that the transfer directions in triadic interactions/relaxation in general depend on the given (relative) amplitudes of the Fourier modes. The simple examples are the cases of the familiar nonhelical initial energy spectra \textit{ansatz} $k^a$ in 3D: if $a>2$, the large-$k$ modes' energies are higher than equipartition (`too hot') and that should be transferred to smaller $k$, contrary to the common forward turbulence cascade, unless that for some special reason spontaneous symmetry breaking, as we just proposed for the findings of Chen et al. \cite{CCEH05} and Bourouiba et al. \cite{BourouibaJFM12}, happens to allow a subsystem with the helical absolute equilibrium to indeed be able to contain the much higher energy at the smallest scales \cite{k73}; similarly for pure 2D case, if the initial energy and enstrophy are respectively too high at the two ends of the wave number range (`too hot' - `hotter' means `more negative' of the negative temperature, because the more negative the temperature is, the larger is the singular wavenumber, thus the higher is the energy level around there), their transfer directions during the relaxation are contrary to the ones usually observed. Now, as a possibility, suppose `initially' (starting from some time the partial decoupling occurs) the temperatures of the system are somehow set up, say, (mainly) by just the $2w-w$ interactions due to the distributions of the invariants and the interactions at that moment (assumed to be in quasi-equilibrium), then for some reason, due to the re-distribution of the invariants and the change of the strength of the coupling, the slow-mode subsystem feels `too hot' (the hotter the more negative is $\Gamma_{\mathcal{E}}$) so that $2w-w$ interactions want to transfer $\mathcal{Z}$ forwardly; but, the coupled waves however may feel the slow-mode subsystem still `too cold' (in the sense of the full Euler absolute equilibrium) and keep warming it by $33-w$ interactions: A steady cycle of the energy may be formed by such imbalance, and such `too hot' and `too cold' dissymmetry appears to be in line with the trigger of the `spontaneous mirror symmetry breaking' of exchanging opposite-sign helicities just mentioned. Of course the system is not in equilibrium, but taking the objective absolute equilibria as the aim the system tends to relax is a good way of thinking. %Note also that, as stated, transfer directions of $2w-w$ indicated by the absolute equilibria are quite uncertain and the genuine transfers should be depending on the actual situations in a sensitively nonuniversal way, thus i
If one imposes large-scale $\mathcal{Z}$ friction, transfers to large scales
of $\mathcal{Z}$ in a statistical steady state might `naturally' emerge.

The conservation of these quantities was shown to be reasonably accurately
valid only in the small Ro regime in Ref. \cite{BourouibaPoF08}, not in the
moderate/intermediate-Ro case. How can our argument work for
moderate/intermediate Ro, and is it possible for the existence of additional
invariant(s) introduced by resonant and near-resonant modes (see, e.g.,
Smith and Lee\cite{SmithLee05} and references therein)? The answer lies in
that we are talking about the persistent effects of the otherwise
self-autonomous 2D3C subdynamics as a conceptual comprehension, not any sort
of precise shape or exact dynamics. This is reminiscent of the finding of
the partial thermalization at the end of the inertial
range\cite{FrischPRL08}, where hyperviscosity or other variants with
appropriate parameterization \cite{ZhuTaylor10} can greatly enhance the
bottleneck phenomena as the residue of thermalization: Although theory says
about the asymptotic infinite parameter (say, the hyperviscosity in Ref.
\cite{FrischPRL08}) limit, the bottleneck appears already in the normal
fluid case and is already greatly strengthened for `moderate/intermediate'
parameters there. On the other hand, such remarks mean that the relevant
argument does not need to consider that complete two-dimensionalization and
separation between fast (waves) and slow (vortex) modes by fast rotation are
unconditionally ascertained in the large-Ro limit. As for the near-resonant
modes (see, e.g., Refs. \cite{SW99,SmithLee05} and references therein), it
may be that they, or at least part of them with small $|k_z|/k$, can be
included into the slow manifold as if they also obey the same conservation
laws and the statistical mechanics, at the appropriate time regime for good
approximation: For example, one may further check whether in the data the
near-resonant modes, kicking in first with $33-2$ and $33-w$ interactions,
are those with small $|k_z|/k$, i.e., the `near-slow' modes whose dynamical
time scales are relatively slow. Of course, another speculation is that the
resonant manifold (including the slow one) may present its own special
conservation laws and statistical mechanics (favoring the inverse
transfers), and that near-resonant modes help the interactions between the
slow and fast modes in this manifold. However, the following analysis shows
that the resonant condition does not introduce extra generic invariant(s):
The already known symmetry relations of the full dynamics in the rotating
frame and the slow-mode dynamics are the same,
$\hat{C}^{c_{\bm{k}}}_{c_{\bm{p}}{c_{\bm{q}}}}+\hat{C}^{c_{\bm{p}}}_{c_{\bm{q}}{c_{\bm{k}}}}+\hat{C}^{c_{\bm{q}}}_{c_{\bm{k}}{c_{\bm{p}}}}=0$
and
$c_{\bm{k}}k\hat{C}^{c_{\bm{k}}}_{c_{\bm{p}}{c_{\bm{q}}}}+c_{\bm{p}}p\hat{C}^{c_{\bm{p}}}_{c_{\bm{q}}{c_{\bm{k}}}}+c_{\bm{q}}q\hat{C}^{c_{\bm{q}}}_{c_{\bm{k}}{c_{\bm{p}}}}=0$,
corresponding to the conservation of energy and helicity respectively, which
can be expressed as\cite{W93}
\begin{equation}\label{eq:WaleffeCoupling}
    \frac{\hat{C}_{c_{\bm{p}}c_{\bm{q}}}^{c_{\bm{k}}}}{c_{\bm{q}}q-c_{\bm{p}}p}=
\frac{\hat{C}_{c_{\bm{q}}c_{\bm{k}}}^{c_{\bm{p}}}}{c_{\bm{k}}k-c_{\bm{q}}q}=\frac{\hat{C}_{c_{\bm{k}}c_{\bm{p}}}^{c_{\bm{q}}}}{c_{\bm{p}}p-c_{\bm{k}}k};
\end{equation}
and similarly
\begin{equation}\label{eq:WaleffeResonant}
    \frac{k_z/k}{c_{\bm{p}}q-c_{\bm{q}}p}=
\frac{p_z/p}{c_{\bm{q}}k-c_{\bm{k}}q}=\frac{q_z/q}{c_{\bm{k}}p-c_{\bm{p}}k}
\end{equation}
for the triadic interaction condition $k_z+p_z+q_z=0$ and the resonant
condition $c_{\bm{k}}k_z/k+c_{\bm{p}}p_z/p+c_{\bm{q}}q_z/q=0$. Using the
fact that
$$c_{\bm{p}}c_{\bm{q}}(c_{\bm{p}}q-c_{\bm{q}}p)=c_{\bm{q}}q-c_{\bm{p}}p$$ in
the comparison between Eqs. (\ref{eq:WaleffeCoupling}) and
(\ref{eq:WaleffeResonant}), we immediately see that \textit{the resonant
condition corresponds just to the already known energy-conservation
relation/symmetry}. So, we can not find new symmetry/conservation law from
the additional information of resonant condition, and the reason appears to
be the simple relation $c_{\cdot}^2=1$ not used in the derivation of the
`old' symmetry relations of $C_{\cdot\cdot}^{\cdot}$. Whether or not the
near-resonant interactions (depending on the time scale of choice
\cite{Newell69}) would introduce new invariant(s), which could be crucial
for the dynamics, is however, to our point of view, so far completely
clueless from the state-of-the-art \cite{SmithLee05}. So, we tend to extend
the following point of Smith and Lee \cite{SmithLee05} to also the
generation of large-scale vertically-averaged vertical velocity: ``an
inverse cascade in the 2D plane alone does not fully explain the generation
of 2D large-scale motions in 3D rotating flows, at least at moderate Rossby
numbers where numerical simulations can adequately resolve near resonances.
Nevertheless, 2D interactions are crucial for the generation of large
scales.'' That is, \textit{the large-scale concentration of $\mathcal{Z}$ we
found in the 2D3C absolute equilibrium, due to the limitation to 2D, might
be far from sufficient to account for Chen et al. \cite{CCEH05} and
Bourouiba et al. \cite{BourouibaJFM12}'s corresponding findings, but may
play a crucial role}. If inverse transfers observed by Chen et al. and
Bourouiba et al. are indeed generic, there must be an intrinsic drive of
this type playing the key role, to our belief. Actually, we also tried to
explain relevant puzzling inverse transfers in rotating flows with another
absolute-equilibrium argument (Appendix), which, though interesting, however
is believed to be not as intimate.

%One can continue posing other phenomenological questions, but we feel that t
%The `question' lying in the core of the problem to `answer' all questions is simply that, if the calculated absolute equilibrium is indeed accurate and represents the relaxation direction of the triadic interactions, why the concentration at large scales of the scalar energy, which is free of any effective constraint, would not induce an inverse transfer in the turbulent state? And coming back to the beginning of the introductory discussion, how can passive scalar transfer be universal, given that the pumping can be chosen to adapt to the advecting field to create correlation? N

\section{Controlling $\mathcal{C}$}\label{sec:ControlC}
\subsection{The (un)controllability of $\mathcal{C}$}
$\mathcal{C}$ can be set up in the initial condition and can have the
corresponding dynamical effects during the decaying process: `everything'
will die out in the end, but the transient process may be affected by
$\mathcal{C}$; for instance, if $\mathcal{C}$ is appropriately configured in
the beginning, then $\mathcal{Z}$ may be transferred to largest scales or be
persistently staying at large scales. For the pumped statistically steady
state, it is intriguing whether we can well control $\mathcal{C}$ and thus
the corresponding dynamics. We see that the external injection of
$\mathcal{C}$ comes from $2(f_{\zeta}\theta+f_{\theta}\zeta)$; and, for a
passive scalar problem, $f_{\zeta}$ is independent of $\theta$, but $\theta$
functionally depends on $f_{\zeta}$; meanwhile, \textit{$f_{\theta}$ can be
taken to be dependent on/correlated to $\zeta$}. The notion and situation of
such \textit{asymmetrical dependence} are not new at all: For example, in
the celebrated Kraichnan model of passive scalar, the scalar pumping is
taken to be independent (of $\theta$) Gaussian white in time, with however
finite correlation $\langle \theta f_{\theta} \rangle=\langle f_{\theta}^2
\rangle/2$ \cite{Novikov}, due to the (functional) dependence of $\theta$ on
$f_{\theta}$. It is helpful to take the Lagrangian point of view and
technique (\cite{FrischVillone14,EyinkJMP09,EyinkNature13} and references
therein). We can integrate along the Lagrangian trajectory
$\bm{R}(\bm{r},t;s)$ that will come to $\bm{r}$ at $t$ (i.e.,
$\bm{R}(\bm{r},t;t)=\bm{r}$) to obtain
$\Theta(\bm{r},t)=\Theta_0(\bm{R}(\bm{r},t;0),0)+\int_0^t
f_{\theta}(\bm{R}(\bm{r},t;s),s)ds$. With the molecular diffusivity treated
as the effect of an added independent `microscopic' uniform Brownian motion,
the average over which should be separable from other randomness and is
denoted by $\overline{\Theta}=\theta$, we rewrite
$\theta(\bm{r},t)=\overline{\int_{-\infty}^t
f_{\theta}(\bm{R}(\bm{r},t;s),s)ds}$ by formally introducing $s=-\infty$,
for notational convenience. Then, taking the zero diffusivity limit and
further averaging over the macroscopic randomness of $\bm{R}$ (or $\bm{v}$)
and $f_{\theta}$ [in general we use the same `$\langle \cdot \rangle$' as in
Eq. (\ref{eq:crosscorrelation}) for statistical average over whatever
statistical ensemble], we have
\begin{eqnarray}
\langle \theta\zeta \rangle=\langle \zeta(\bm{r},t)\int_{-\infty}^t f_{\theta}(\bm{R}(\bm{r},t;s),s) ds \rangle=\langle \int_{-\infty}^t f_{\zeta}(\bm{R}(\bm{r},t;s),s) ds \int_{-\infty}^t f_{\theta}(\bm{R}(\bm{r},t;s),s) ds \rangle,\label{eq:LagrangianC}\\
\langle f_{\zeta}\theta+f_{\theta}\zeta \rangle=\langle f_{\zeta} \int_{-\infty}^t f_{\theta}(\bm{R}(s),s)ds +  f_{\theta} \int_{-\infty}^t f_{\zeta}(\bm{R}(s),s)ds \rangle.\label{eq:injection}
\end{eqnarray}
Here, taking the $\kappa \to 0$ limit does not necessarily remove the
average over the paths: Even for a given realization of velocity, the
statistical average over paths are in general still needed when
spontaneous/intrinsic stochasticity happens with rough velocity, and the
average over the paths and that over the pumping are not necessarily
separable due to possible nontrivial dependence of $f_{\theta}$ on $\bm{v}$
or its derivable(s), the above mentioned $\zeta$, say. Such a stochastic
equation issue makes the problem subtle. For example, in the turbulent state
with $f_{\theta}$ and $f_{\zeta}$ coincident, and even $\nu=\kappa$ and the
same initial and boundary conditions, $\theta=\zeta$ may not hold in the
classical sense (though $\zeta$ of course formally solves the $\theta$
equation \cite{ConstantinMajdaCMP88}), due to the rough velocity field. The
inverse energy cascade range of 2D turbulence with a spectrum of exponent
$-5/3$ corresponds to a rough velociby field, and in the forward enstrophy
cascade, logarithmic correction also may introduce weak stochasticity of
trajectories. Such nonuniqueness or stochasticity of the field requires a
probabilistic description. So, we can not see that taking
$f_{\theta}=f_{\zeta}$ (coincident) necessarily optimize $\mathcal{C}$.
Commonly-accepted forward cascade of $\mathcal{W}$ means that the whole
nonlinearity does not conserve $\mathcal{W}$ for well developed turbulence
(as in the conjecture of Onsager for the kinetic energy in 3D
\cite{EyinkEuler08}). And, even with $\nu=\kappa$ and
$f_{\theta}=f_{\zeta}$, $\langle (\theta-\zeta)^2 \rangle$ would not
necessarily only be damped by the molecular diffusivity/viscosity.
Furthermore, even if both $\theta$ and $\zeta$ are subjected to anomalous
dissipation, their dissipation rates are not necessarily the same when
$\nu=\kappa \to 0$; that is, the (perturbative) difference $\theta-\zeta$,
from the boundary and/or initial conditions or emerging spontaneously, can
persist or even amplify. Turbulent coincidence $\zeta=\theta$, if possible,
must come with other extra strong constraint(s) of the dynamics which might
directly prohibit inverse transfer and is beyond this analysis.
[\textit{Actually we can not even see from the absolute equilibrium whether
the imposition of $\zeta=\theta$ would maximize the condensation of
$\theta$, i.e., whether more of $\mathcal{Z}$ will go to largest scales
while $|\mathcal{C}|$ is increased with fixed $\mathcal{Z}$, $\mathcal{E}$
and $\mathcal{W}$.} Due to the nonlinear structure of Eqs. (\ref{eq:2D3CU}
and \ref{eq:2D3CQ}), it is possible that no such monotonicity exists.]
\textit{A set of numerical experiments may be proposed as follows}: Let the
initial $\theta$ and $\zeta$ fields be different but correlated, with
various values of $\mathcal{C}$, and let the other things as mentioned be
the same, in particular the same forcing working at some intermediate
scales, then check the evolutions of the statistics of $\theta$, $\zeta$ and
their differences $\theta-\zeta$ at different scales.
One however may say that, in Ref. \cite{CCMVnjp04}, if all the corresponding advecting velocity fields/trajectories and passive-scalar forces effectively have the same/similar dynamical correlations respectively, all those passive scalars could have the same/similar \textit{statistics} of the corresponding active ones, including the inverse transfer of the potential energy of 2D MHD: \textit{Such a statement may be numerically checked by using the information of the correlation between the scalar pumping and the tracer trajectories, say, somehow learned from the 2D MHD data, in the passive scalar pumping scheme.} The passive scalar evolve in this way can be quite different to the corresponding 2D MHD active scalar beyond the second-order correlation between the forcing and the trajectories. %What exactly are the characteristics of the pumping and the advecting velocity fields and how they uniquely specify the statistics of the passive scalar, if indeed, however are not clear.
When studying 2D MHD, Ref. \cite{CCMVnjp04} has derived a set of necessary
conditions for the inverse and/or forward cascades of the scalars. Then, it
would be interesting to experiment numerically whether the correlations
between the scalar pumping and the advecting velocity are also sufficient
conditions, especially when the velocity field is not particularly generated
by 2D MHD.

%As said, forward cascade of $\mathcal{Z}$ to small scales (say, together with $\mathcal{W}$) does not by itself exclude the possibility of simultaneous genuine inverse transfer together with $\mathcal{E}$ at large scales (with the pumping scale in between). As can be seen from Eq. (\ref{eq:injection}), the injection of $\mathcal{C}$ is not easily seen to be trivially under precise control.
%:
As said, the passive scalar should be functionally dependent on the stirring of the velocity, besides on its own pumping, so $\langle f_{\zeta}\theta \rangle$ is in general not controllable: \textit{How to optimize the injection is unclear.} %[In usual statistical theory, the forcing is taken to be delta-correlated in time, which however may not be optimal to inject $\mathcal{C}$, for the response of $\theta$ to $f_{\zeta}$ through $\bm{v}$ can have time delay, for instance.]
But %Writing $\langle f_{\omega}\theta \rangle=\langle f_{\omega}\int_{-\infty}^t f(\bm{R}(s),s)ds \rangle$, we see that
$\langle f_{\zeta}\theta \rangle$ \textit{could} be small and is, for the
time being, \textit{assumed} so here for a tentative discussion of modifying
the Kraichnan model. So, we may take the velocity be a synthetic
delta-correlated Gaussian field and let $f_{\theta}$ be linearly correlated
to $\zeta$, thus still Gaussian, to inject $\mathcal{C}$.
%- though how $\langle f_{\omega}\theta \rangle$ is changed by such a particular forcing is unknown, we might assume the change should be little by the independence between the pumping and forcing. Anyhow, among various uncertainties, it still appears that we may try to control the injection of $\mathcal{C}$ by such correlation of $f_{\theta}$ to $\omega$, and it is the only possibility we have found so far.
%It is interesting to n
Note that such a correlation is due to the dependence of $f_{\theta}$ on
$\zeta$, not the inverse, which is not represented by the correlation
itself.
%Another way to look at the problem is by writing $\langle f_{\omega}\theta \rangle=\langle f_{\omega}\int_{-\infty}^t f(\bm{R}(s),s)ds \rangle$ which indeed should vanish.
The purpose of the modification is to see whether we can find a way out of
the conventional Kraichnan model's universal (in the sense of the scaling
exponents with respect to the pumping mechanisms) forward cascade of
$\mathcal{Z}$. We iterate that we are proposing \textit{passive} correlation
of the pumping on $\zeta$, which is different to the probability-theoretic
correlation which is symmetric between the two random variables. In general
the correlation will make the average, over the pumping ensemble, and the
other one, over the velocity ensemble, mixed, while the decoupling of the
two averages are crucial for the systematic calculations \cite{FGVrmp01};
but, carefully designed coupling in some specific situation might be
insightful.

\subsection{Pair-correlation function and an attempt to modify the Kraichnan model}
We now check the integral form of the pair-correlation function
\begin{equation}\label{eq:pair-correlationFunction}
    C_2(\bm{r}_1,\bm{r}_2;t)=\langle \theta_1 \theta_2 \rangle = \Big{\langle} \int_{-\infty}^{t} f_{\theta}(\bm{R}(\bm{r}_1,t;s_1),s_1)ds_1 \int_{-\infty}^{t} f_{\theta}(\bm{R}(\bm{r}_2,t;s_2),s_2)ds_2 \Big{\rangle}.
\end{equation}
%The linear dependence of the pumping on the vorticity does not uniquely specify the statistics of the velocity field.
%especially the spacial correlation of the pumping (presumably the same as that of the vorticity field, but we don't know how to further average the two-point correlation over the velocity field) here, which appears to be the corresponding problem of the above.
%If the pumping, like the velocity, were independent stationary, Gaussian, homogeneous, isotropic, of zero mean and with covariance $\langle f(\bm{r},t)f(\bm{0},0) \rangle=\delta(t)\Phi(r)$, and even if we could (generally not) decouple the average over pumping and that over velocity, then we would have $C_2(\bm{r}_{12},t)=\Big{\langle} \int_{-\infty}^t \Phi(R_{12}(s))ds \Big{\rangle}$ (see, e.g., \cite{FGVrmp01}), where $\bm{R}_{12}$ indicates the separation between the two positions. %and where $\Phi(r)$ is the two point correlation of the pumping.
But, without knowing how $f$ depends on $\bm{v}$, still we don't know how to proceed with the average, also denoted by $\langle \cdot \rangle$ here, over the velocity ensemble. %It is obvious that d
Different dependences of $f$ on $\bm{v}$ will lead to different results.
%[For the single point average, $\langle \theta \rangle= \Big{\langle} \int_{-\infty}^{t} f[\bm{R}(s),s;\bm{r},t]ds \Big{\rangle}$ can be determined because of the local linear relation between the velocity and vorticity fields.]
%The `failure' to do the calculations to the end does not mean that we are completely blocked from enlightening: It is probably exactly such a `failure' or `uncertainty' of calculations that underlies the `nonuniversality'.
So, let us further perform more explicit exposition, % in the following two subsections,
with the hope of finding a way out in some special situations: There could
be some deliberately designed statistical dependence which is solvable and
useful.
%: Different types of the gauge field $\bm{g}(\bm{x})$ may correspond to different statistical dynamics, thus it would be interesting to specify some examples as we will try to show.
%\textit{Of course, if the pumping was taken linearly dependent directly on the advecting velocity field (as in, e.g., Holzer and Siggia \cite{HolzerSiggia94} with $f_{\theta}=\bm{v}\cdot \bm{G}$ with $\bm{G}$ being the constant mean gradient of the scalar), then the averages over pumping and velocity would be the same and could be taken simultaneously and that there should be no problem to carry out the systematic calculations (though `useless', for no $\mathcal{C}$): For example, Eq. (135) of Falkovich et al. \cite{FGVrmp01} is now understood to be the average over both the pumping and velocity simultaneously.}
Note that our purpose does not need to be as ambitious as to fully solve the
problem as for the conventional Kraichnan model, but to draw the information
of transfer directions, or more definitely the possibility of inverse
transfer at large scales when $\mathcal{C}$ is appropriately injected and
sustained.
%\subsection{First order moment equation}
Here we adopt the pedagogical functional differential approach of Frisch and
Wirth \cite{FrischWirthLNP97} for just a preliminary analysis. When the
pumping $f_{\theta}$ is linearly correlated to the vorticity, i.e.,
$f_{\theta}=\nabla\times\bm{v}\cdot\bm{z}$ (the arbitrary coefficient is
normalized to unit), with the assumption that the perturbation of the
advection operator is due to the perturbation of the advecting velocity, Eq.
(\ref{eq:p1}) becomes
\begin{eqnarray}
% \nonumber to remove numbering (before each equation)
  \partial_t \theta + \bm{v}\cdot\nabla\theta&=& \kappa\nabla^2\theta+\nabla\times\bm{v}\cdot\bm{z}. \label{eq:p3a}
\end{eqnarray}
Now, $\bm{v}(\bm{x},t)$ is Gaussian white in time, so, with $\partial_{\alpha}\theta|_{\alpha=0}$ denoted by $\underline{\theta}(\bm{x},t)$, we have %[c.f., Eq. (10) of Ref. \cite{FrischWirthLNP97}]
\begin{eqnarray}
% \nonumber to remove numbering (before each equation)
  \partial_t \underline{\theta} = M\underline{\theta}+\tilde{M}(\bm{x},t) \underline{\theta}+\tilde{M}'(\bm{x},t) \theta - \nabla\times\bm{v}'\cdot \bm{z} , \label{eq:palpha1}
\end{eqnarray}
where $M=\kappa\nabla^2$ and $\tilde{M}=-\bm{v}\cdot\nabla$ and $'$ denotes
``identical independent distribution (i.i.d.)'' Note that the perturbation
$\tilde{M}\to \tilde{M}+\alpha \tilde{M}'$ means that $-\bm{v}\cdot\nabla
\to -(\bm{v}+\alpha \bm{v}')\cdot\nabla$. So, we have
\begin{eqnarray}
% \nonumber to remove numbering (before each equation)
\underline{\theta} = \int_{t_0}^t -G(t,s)[\bm{v}'(s)\cdot\nabla\theta(s) +\nabla\times\bm{v}'(s)\cdot\bm{z} ] ds, \label{eq:palpha2}
\end{eqnarray}
with $\underline{\theta}(t_0)=0$ and $G$ being the Green's function
corresponding to the operator $M+\tilde{M}$, and that we have the averaged
equation
%[c.f., Eq. (6) of Ref. \cite{FrischWirthLNP97}]
\begin{eqnarray}
\partial_t \langle \theta \rangle=\kappa\nabla^2 \langle \theta \rangle + \int_{t_0}^t \langle \tilde{M}'(t)G(t,s)[\tilde{M}'(s)\theta(s)-\nabla\times\bm{v}'(s)\cdot \bm{z}]\rangle ds.\label{eq:m1}
\end{eqnarray}
This equation is closed and in principle solvable when the condition of
white in time is applied, provided the information of the spacial derivative
of the velocity field.
%\subsection{Second order moment equation}
To study the second-order moment equation, we use further simplified
notation $f_{\theta}=\nabla\times\bm{v}\cdot\bm{z} \to f$ and
$\theta(\bm{x}_i) \to \theta_i$ for $i=1,2$ etc. Two-point equation reads
\begin{eqnarray}
\partial_t (\theta_1\theta_2)+(\bm{v}_1\cdot\nabla_1+\bm{v}_2\cdot\nabla_2)(\theta_1\theta_2)=\kappa(\nabla_1^2+\nabla_2^2)(\theta_1\theta_2)+f_1\theta_2+f_2\theta_1.\label{eq:m2}
\end{eqnarray}
Among other things, to obtain $\langle \theta_1\theta_2 \rangle$ from the
above equation, we need to evaluate in the r.h.s. $\langle
f_1\theta_2+f_2\theta_1 \rangle$, by Gaussian integration by parts: $\langle
f_1\theta_2\rangle=\langle f'_1\underline{\tilde{\theta_2}} \rangle$. Note
that $\underline{\tilde{\theta_2}}$, with the perturbation directly from $f$
(thus the tilde for symbolic discrimination), is not that of Eq.
(\ref{eq:palpha2}), since it does not obey Eq. (\ref{eq:palpha1}): The
perturbation of the pumping, linearly dependent on the vorticity, does not
uniquely determine the perturbation of the advection velocity/operator,
leaving the freedom of the gauge field $\bm{g}(\bm{x},t)$ with
$\nabla\times\bm{g}=0$. And, it is seen that $\underline{\tilde{\theta_2}}$
is not $\bm{g}$-free. So, in this sense, the second order moment equation
with the pumping linearly correlated to the vorticity turns out to be
\textit{intrinsically} non-unique.
%[unlike Eq. (36) of Ref. \cite{FrischWirthLNP97} for the standard Kraichnan model where $f$ and $\bm{v}$ are independent]!
[The systematic calculations of the standard Kraichnan model can be made to conclude the universal (in the sense of independence on the pumping mechanisms of the scaling exponents) forward cascade of the passive scalar, because the $f$ and $\bm{v}$ are mutually independent Gaussian white in time and also that $f$ is chosen at some large scales.% Statistical dependence between $f$ and $\bm{v}$ hinders the procedure of separating the average over $f$ and that over $\bm{v}$.
]
Arbitrarily different gauge fields of $\bm{g}(\bm{x})$ may correspond to different underlying `details' of the pumping, resulting in different statistical dynamics with strong indication of assisting the nonuniversality. %, thus it would be interesting to specify some examples as we will try to show.
%$\bm{g}(\bm{x})$ can only be fixed by extra constraint(s).
Though such a result is in a sense already foreseeable from the beginning,
the above preliminary analysis explicitly exposes how uncontrollable the
closure for the problem is. As is clear from, say, Falkovich et al.
\cite{FGVrmp01} and Kramer et al. \cite{KramerMajdaVanden-Eijnden03}, one
should be careful in extrapolating results from the celebrated Kraichnan
model of passive scalar turbulence and any theory with \textit{ad hoc}
assumptions, however complicated it is (like, say, `DIA' \cite{k68}), can be
very misleading in some cases \cite{KramerMajdaVanden-Eijnden03}; thus, a
systematically solvable model with controllable $\mathcal{C}$ is in fact
still wanted.

%\section*{\textbf{\S V}
\section{On `vertical helicity' and the speculation of 2D3C dominated helical
cyclogenesis}\label{sec:Remark}
%\begin{CJK*}{GB}{}
Finally, for both fundamental and application reasons, we want to examine
other systems. Indeed, like 2D vorticity there are different `frozen in'
quantities corresponding to Lagrangian conservation laws [by, say,
Kelvin(-type) or Alfv\'en theorems, or their analogies] for a variety of
fluid models of plasmas (see, e.g., Ref. \cite{EyinkJMP09} and references
therein), however it remains to discover one whose multiplication with the
auxiliary passive scalar can form a rugged invariant. One possible way may
be to extend the notions of vorticity and helicity to $n>3$ dimensions with
the notion of differential forms and then reduce the problem to $n-1$
dimensional space with still $n$-component vectors. Such formal studies
won't be pursued in this note.
Below we will just point out another specific atmosphere application consideration; see the Appendix for more. In (hazardous) weather forecasting or investigations, the so-called ``vertical helicity'' $u_z\zeta$ is frequently applied (see, e.g., Kain et al. \cite{Kain08}). In our terminology, as is also clarified in many other literatures, $u_z\zeta$ is actually the `vertical helicity \textit{density}'. It is convenient to simplify and clarify such weather-science terminology `helicity (density)' %[`???(???)']
by `\textit{helixity}', thus `vertical helixity', `horizontal helixity' etc.
Vertical helixity is also used in studying sand- and/or dust-storms. We see
from our 2D3C analysis that this vertical helixity is paired with the
horizontal helixity $$-\bm{z}\times\nabla u_z \cdot \bm{u}_h,$$ with equal
spacial integrals.  Note that the full horizontal helixity contains also
other parts which vanish with $\partial_z=0$. It may thus be suggestive to
even further re-define the vertical helixity to be just the density of
$\mathcal{C}$, $$u_z \zeta -\bm{z}\times\nabla u_z \cdot \bm{u}_h,$$ or at
least to also examine the weather data with the `conjugate'
$-\bm{z}\times\nabla u_z \cdot \bm{u}_h$ in the pair. And, from the possible
large-scale formation and amplification of both $\mathcal{Z}$ and
$\mathcal{C}$ [Eqs. (\ref{eq:2D3CU}) and (\ref{eq:2D3CQ})], we can not
resist speculating \textit{2D3C dominated helical cyclogenesis} (and
large-scale smog/haze formation and strengthening/persistence, if the
pollutant could be treated as passive scalar). The idea of helical
cyclogenesis actually has a long history, to our best knowledge, dated back
at least to Levich and Tzvetkov \cite{footnoteLevichTzvetkov84} (see, Levina
and Montgomery \cite{LevinaMontgomery14} and references therein for extra
relevant information) whose discussions deserve some remarks here, though
their arguments of `$I$-invariant' and its inverse cascade appear to us
somewhat strange: To our understanding, if the spacial average and
statistical average are taken to be equal, their $I$ is nothing but the
square of the conventional global invariant helicity. $I$ is not conserved
in detail by the triadic interactions, thus not rugged concerning Galerkin
truncation, and that not a good object for talking about `cascade'.
Nevertheless, they did explicitly use the 2D3C argument [Eq. (9) there] and
did argue non-Kraichnan \cite{k67} inverse cascade, though not for our
$\mathcal{Z}$ and or $\mathcal{C}$. Their inverse-cascade scenario for
cyclogenesis, assisted with the meteological figures of cyclones, clouds and
precipitation, is similar, at least on the surface, to our speculation,
concerning the relevance of helicity.

\section*{Appendix}
Here we summarize two absolute-equilibrium related theories for inverse
transfers observed in rotating flows. One indicates extra properties of
nonlocality of interactions, while the other clearly identifies itself in
the slow manifold as given in the main text. The former one indicates
strikingly similar turbulence energetic behavior to reported numerical
observations, but the corresponding Galerkin truncation appears extremely
artificial, though interesting and might be of other relevance. It was meant
to be added to Ref. \cite{hydrochirality} at its late stage (but was not
allowed due to the acceptance of publication by the journal at the time), as
our two theories motivated by the same/similar numerical observations of
rotating flows\cite{SW99,map09,MP10}. We now consider this old discussion be
still useful to offer extra insights about helicity, with \textit{new}
motivating perspectives. It was attached to Ref. \cite{hydrochirality}, so
we will briefly cite a result that we need here first: \begin{figure}
          % Requires \usepackage{graphicx}
          \begin{center}
          \includegraphics[width=0.75\textwidth]{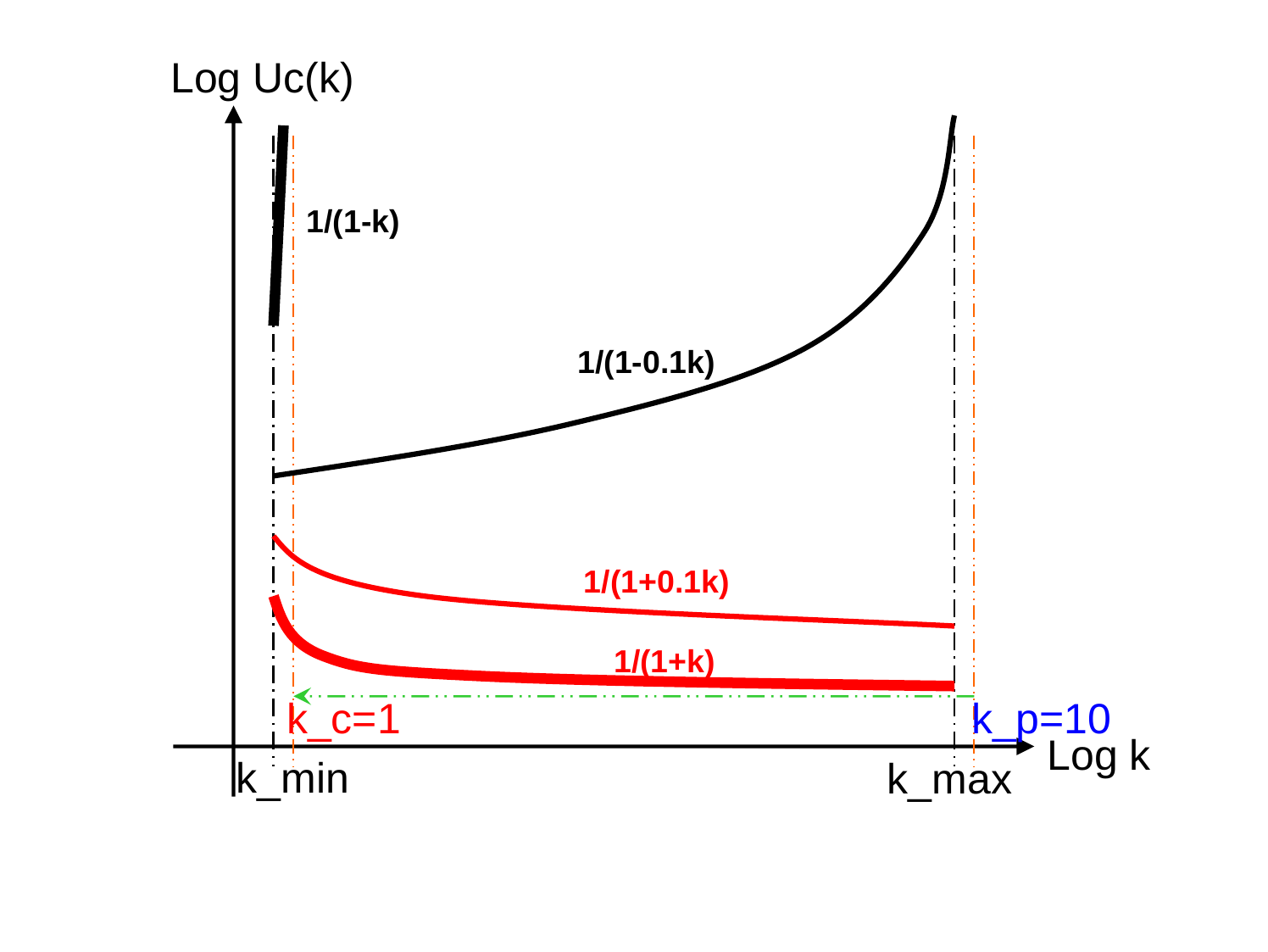}\\
          \caption{Schematic for $U^c(k)=1/(\alpha+c\beta k)$: $0<k_{min}\lesssim k_c=1$ and $k_{max}\gg k_c$. When $\alpha/\beta=1/0.1$ (thinner lines), the negative chiral sector pole $k_p=10\gtrsim k_{max}$. This pole approaches $k_c$ when $\alpha/\beta \to 1/1$ (thicker lines) and most of the energy is concentrated at the negative chiral sector around the pole $k_p \to k_c=1$, the very small wavenumber compared to $k_{max}$: $k_c$ is the maximum wavenumber of the positive chiral sector, so the (dominant) positive helicity has opposite sign to the larger-wavenumber modes where injection is supposed be placed in the turbulence simulations of Refs. \cite{map09,MP10}.}\label{fig:ocsds}
          \end{center}
        \end{figure}
As sketched in Fig. \ref{fig:ocsds}, we found from the chirally decomposed
energy absolute equilibrium spectrum of the full 3D3C system, $U^c(k)
=1/(\alpha+c\beta k)$, that, if one chiral sector (say, that of `$c=+$' for
purely helical modes with positive definite helicity) is truncated to be in
between $k_{min}$ and $k_{max}$, and the other `alien' sector in between
$k_{min}$ and $k_c$ ($\ll k_{max}$), then the `alien(s)' can carry most of
the energy when $\alpha/\beta \downarrow k_c^+$ (with the pole $k_p$
approaching $k_c$ from above), especially when $k_c$ is close to $k_{min}$.
Of course, this result also applies to the system in a rotating frame. And,
more fascinatingly, if we keep the positive and negative sectors in this
fashion for a turbulence system with dissipation at large $k$ and injection
at some $k_{in}$ inbetween $k_{min}$ and $k_{max}$, then the indication for
large scale behavior and nonlocal transfers (see details in Ref.
\cite{hydrochirality}) is very similar to Mininni et al.'s observations
\cite{map09,MP10}, except for one thing (see below), at least.

To be more definite, Mininni at al.\cite{map09} stated ``... the direct
transfer of energy at small scales is mediated by interactions with the
largest scale in the system, the energy containing eddies with
$k_{\perp}\approx 1$, where $k_{\perp}$ refers to wavevectors perpendicular
the axis of rotation. ... The inverse transfer of energy at scales larger
than the energy injection scale is non-local, and energy is transferred
directly from small scales to the largest available scale'', \textit{which
are basically the same as our prediction from the special truncation scheme
in \S 2.2.2 of Ref. \cite{hydrochirality}, word by word (if we appropriately
choose the directions of the wavevector and the corresponding helical mode
of the alien(s))}; what's more, Mininni and Pouquet\cite{MP10} found the
``the development of an inverse transfer of energy now coexisting with
direct cascades of energy and helicity''. One probably then would guess that
the imposed rotation work to naturally depress those modes that are cleanly
truncated in the special scheme, which however appears somewhat strange.
More seriously, there is no evidence from these simulations that the largest
scale is `OCSDS' (one chiral sector dominated state) with chirality opposite
to the input (as predicted from the above special absolute equilibrium). We
then believed that the inverse energy transfer in rotating turbulence
probably had not much to do with the helical energy condensations found in
\S 2.2.2 of Ref. \cite{hydrochirality}; in other words, the absolute
equilibrium sketched in Fig. \ref{fig:ocsds} should not be as relevant to
rotating flows as that in Fig. \ref{fig:UhUz}. It was nevertheless hoped
that such analysis could promote relevant numerical analysis in the fashion
of Brandenburg et al. \cite{BrandenburgETC} and/or Biferale et
al.\cite{bmt12},
%(see later even rigorous mathematical estimation \cite{BiferaleTitiJST13}),
but looking into the contributions to the spectral transfers functions from
the purely helical modes.

Thus, we turned to deal with three separated subsystems, {\it i.e.}, the 2D
$u_x$-$u_y$ field (S2) with $\partial_z\equiv 0$, which conserves energy and
enstrophy, the passive scalar $u_z$ field (S1 with scalar energy conserved)
with $\partial_z=0$ advected by S2, and the left 3D field (S3, conserving
energy and helicity) with $\partial_z$ not vanishing everywhere. We can have
the normal energy-helicity AE (absolute equilibrium) of S3 indicating
forward energy and helicity cascades; S2 is already discussed by
Kraichnan\cite{k67}, indicating inverse-energy and forward-enstrophy
cascades; S1 by itself, without coupling to S2, presents equipartition of
the AE energy, indicating forward energy cascade. Note that $S1\oplus S2$,
the combination, also conserves helicity on the slow-mode manifold,
facilitating energy-enstrophy-helicity AE. The indications of S2 and S3 AEs
are consistent with the \textit{two-dimensionalization} and
forward-energy-and-helicity-cascade scenarios as all simulations and
observations seem to agree. 2D-3D \textit{vortex-wave interaction} may be
termed \textit{re-coupling} of S2 and S3, which raises the complex issue of
the validity of time regime of the resonant wave theory, for given $Ro$,
among other relaxation time scale problems. When resonant wave theory breaks
down, we may consider $S1\oplus S2 \oplus S3$, {\it i.e.}, the whole system,
energy-helicity AE, which then raises the curious question about the
difference between such a state and those of non-rotating ones in \S 2.2 of
Ref. \cite{hydrochirality}. Thus the problem can be extremely complicated
depending on the physical parameters ($Ro$ and $Re$ etc.) and time regimes.
Such remarks about absolute equilibria of subsystems of rotating fluid
models may also apply to those of 3D gyrokinetics, among others, with fast
wave dynamics and slow modes such as zonal flows in nature (say, planet
Jupiter) or laboratory such as the tokamak \cite{ZhuHammettPoF10}. Indeed,
(partial) anisotropization is common to many typical systems and unified
treatments, as tried by Cambon and Godeferd \cite{CambonGodeferd93}, may be
possible.


\begin{thebibliography}{99}

%%%%%%%%%%%%%%%%%%%%%%%%%%%%%%%%%%%%%%%%%%%%%%%%%%%%%%%%%%%%%%
%%%%%%%%%%%%%%%%%%%%%%%%%%%%%%%%%%%%%%%%%%%%%%%%%%%%%%%%%%%%%%

\bibitem{FavierETCjfm11} B. Favier, F. S. Godeferd, C. Cambon, A. Delache
    and
    W. J. T. Bos, Quasi-static magnetohydrodynamic turbulence at high
    Reynolds number. J. Fluid Mech. {\bf 681}, 434 (2011).


\bibitem{Cambon90} C. Cambon, Homogeneous MHD turbulence at weak magnetic
    Reynolds numbers: approach to angular-dependent spectra. In Advances in
    Turbulence Studies: Progress in Astronautics and Aeronautics (ed. H.
    Branover \& Y. Unger), vol. 149, pp. 131-145. AIAA(1990).

\bibitem{Moffatt67} H. K. Moffatt, On the suppression of turbulence by a
    uniform magnetic field. J. Fluid Mech. {\bf 28}, 571-592 (1967).

\bibitem{SmithChasnovWaleffe96} L. M. Smith, J. L. Chasnov and F. Waleffe,
    Crossover from Two- to Three-Dimensional Turbulence. Phys. Rev. Lett.
    {\bf 77}, 2467 (1996).



\bibitem%[Celani, Musacchio \& Vincenzi(2010)]
{CMVprl10} A. Celani, S. Musacchio and D. Vincenzi, Turbulence in More than
Two and Less than Three Dimensions. Phys. Rev. Lett. {\bf 104}, 184506
(2010).


\bibitem%[Montgomery \& Turner(1982)]
{MontgomeryTurnerPoF81} D. Montgomery \& L. Turner, Anisotropic
magnetohydrodynamic turbulence in a strong external magnetic field. Phys.
Fluids {\bf 24}, 825 (1981).

\bibitem{M69} H. K. Moffatt, The degree of knottedness of tangled vortex
    lines. J. Fluid Mech. {\bf 35}, 117-129 (1969).


\bibitem%[Kraichnan(1973)]
{k73} R. H. Kraichnan, Helical turbulence and absolute equilibrium. {\em J.
Fluid Mech.\/} {\bf 59}, 745--752 (1973).

\bibitem%[Celani et al.(2004)]
{CCMVnjp04} A. Celani, M. Cencini, A. Mazzino and M. Vergassola, Active and
passive fields face to face. New Journal of Physics {\bf 6} 00 (2004).

\bibitem%[Falkovich, Gaw\c{e}dzki \& Vergassola(2001)]
{FGVrmp01} G.  Falkovich, K. Gaw\c{e}dzki \& M. Vergasolla, Particles and
fields in fluid turbulence. Rev. Mod. Phys. {\bf 73}, 913--975 (2001).

\bibitem%[Kraichnan(1968)]
{k68} R. H. Kraichnan, Small-Scale Structure of a Scalar Field Convected by
Turbulence. Phys. Fluids {\bf 11}, 945--953 (1968).

\bibitem%[Fyfe \& Montgomery (1976)]
{FyfeMontgomery76} D. Fyfe and D. Montgomery, Inverse cascade of magnetic
potential in 2D magnetohydrodynamics. J. Plasma Phys. 16, 181 (1976).

\bibitem%[Holzer \& Siggia(1994)]
{HolzerSiggia94} M. Holzer \& E. D. Siggia, Turbulent mixing of a passive
scalar. Phys. Fluids {\bf 6}, 1820--1837 (1994).


\bibitem{Newell69} A. Newell, Rossby wave packet interactions. J. Fluid
    Mech. {\bf 35}, 255-271 (1969).


\bibitem%[Chen et al.(2005)]
{CCEH05} Q. N. Chen, S. Chen, G. L. Eyink \& D. Holm, Resonant interactions
in rotating homogeneous three-dimensional turbulence. {\em J. Fluid Mech.\/}
{\bf 542}, 139--163 (2005).

\bibitem{CambonRubinsteinGodeferd04} C. Cambon, R. Rubinstein and F. S.
    Godeferd, Advances in wave turbulence: rapidly rotating flows. New J.
    Phys. {\bf 6}, 73 (2004).

\bibitem{SmithLee05} L. M. Smith and Y. Lee, On near resonances and symmetry
    breaking in forced rotating flows at moderate Rossby number. J. Fluid
    Mech. {\bf 535}, 111 (2005).


\bibitem{CambonJacquinJFM89} C. Cambon and L. Jacquin, Spectral approach to
    non-isotropic turbulence subjected to rotation. J. Fluid. Mech.
    \textbf{202}, 295-317 (1989).

\bibitem{CambonMansourGodeferdJFM97} C. Cambon, N. N. Mansour and F. S.
    Godeferd, Energy transfer in rotating turbulence. J. Fluid. Mech.
    \textbf{337}, 303 - 332 (1997).


\bibitem{BourouibaBartelloJFM07} L. Bourouiba and P. Bartello, The
    intermediate Rossby number range and two-dimensional?Cthree-dimensional
    transfers in rotating decaying homogeneous turbulence. J. Fluid Mech.
    \textbf{587}, 139 - 161 (2007).


\bibitem{BourouibaPRE08} L. Bourouiba, Discreteness and resolution effects
    in rapidly rotating turbulence. Phys. Rev. E {\bf 78}, 056309 (2008).

\bibitem%[Bourouiba et al.(2012)]
{BourouibaJFM12} L. Bourouiba, D. N. Strauba, M. L. Waite, Non-local energy
transfers in rotating turbulence at intermediate Rossby number. {\em J.
Fluid Mech.\/} {\bf 690}, 129--147 (2012).

\bibitem{SW99} L. Smith and F. Waleffe, Transfer of energy to
    two-dimensional large scales in forced, rotating three-dimensional
    turbulence. Phys. Fluids \textbf{11}, 1608 - 1622 (1999).

\bibitem%[Zhu, Yang \& Zhu(2014)]
{hydrochirality} J.-Z. Zhu, W. Yang \& G.-Y. Zhu, Purely helical absolute
equilibria and chirality of (magneto)fluid turbulence.
{\em J. Fluid. Mech.} {\bf 739}, 479--501 (2014).%; see also: arXiv:1303.3823 [nlin.CD].

\bibitem{map09} P. D. Mininni, A. Alexakis, and A. Pouquet, Scale
    interactions and scaling laws in rotating flows at moderate Rossby
    numbers and large Reynolds numbers. Phys. Fluids 21, 015108 (2009)
\bibitem{MP10} P. D. Mininni and A. Pouquet, Rotating helical turbulence.
    Part I. Global evolution and spectral behavior. Phys. Fluids 22, 035105
    (2010).


\bibitem%[Chandresekhar(1961)]
{c61}
S. Chandrasekhar, %1961
Hydrodynamic and hydromagnetic stability. Oxford University Press (1961).


\bibitem%[Lee(1952)]
{Lee1952} T.-D. Lee, On some statistical properties of hydrodynamic and
hydromagnetic fields. {\em Q. Appl. Math.\/} {\bf 10}, 69--74 (1952).

\bibitem{Hopf52} E. Hopf, Statistical hydromechanics and functional
    calculus, Ratl. Mech. Anal. {\bf 1}, 87-123 (1952).


\bibitem%[Moffatt(2014)]
{Moffatt14} H. K. Moffatt, Note on the triad interactions of homogeneous
turbulence.
{\em J. Fluid Mech.\/} {\bf 741}, R3 (2014).%: This reference presented nice analyses of the single-triad system (STS), in particular comparisons between `triad solutions' and `full-Euler' solutions and the 2D3C/passive scalar characteristics of the 2D3C STS appearing close to our study, but the detailed conservation laws of energy and helicity among the interactions of each triad or of the general Galerkin truncation are not limitted to the STS [L. Onsager, Statistical Hydrodynamics. Nuovo Cimento, Suppl. 6, 279 (1949)].



\bibitem%[Orszag(1977)]
{Orszag}
S. A. Orszag, %1977
Statistical Theory of Turbulence, in Fluid Dynamics, Les Houches 1973,
237-374, eds. R. Balian \& J.L. Peube. Gordon and Breach, New York (1977).

\bibitem%[Lee (1979)]
{JonLee79} J. Lee, Dynamical behavior of the fundamental triadic-interaction
system in three dimensional homogeneous turbulence. Phys. Fluids. {\bf 22},
40--53 (1979).

\bibitem{sphmRapids} J.-Z. Zhu, Generic 3D solenoidal modes with vanishing
    quadratic/nonlinear terms, submitted for publication (2015).

\bibitem%[Kraichnan(1967)]
{k67} R. H. Kraichnan, Inertial ranges in two-dimensional turbulence. {\em
Phys. Fluids\/} {\bf 102}, 1417--1423 (1967).


\bibitem%[Frisch et al.(2012)]
{FrischETCprl12} U. Frisch, A. Pomyalov, I. Procaccia and S. Ray, Turbulence
in noninteger dimensions by fractal Fourier decimation. Phys. Rev. Lett.
{\bf 108} 074501 (2012)

\bibitem%[Zhu(2014)]
{ZhuPoF14} J.-Z. Zhu, Note on specific chiral ensembles of statistical
hydrodynamics: ``Order function'' for transition of turbulence transfer
scenarios. Phys. Fluids {\bf 26}, 055109 (2014).

\bibitem{SheJackson} Z.-S. She and E. Jackson, Constrained Euler System for
    Navier-Stokes Turbulence. Phys. Rev. Lett. {\bf 70}, 1255 (1993).

\bibitem{nonir} G. Gallavotti, Nonequilibirum and irreversibility. Springer
    International Publishing (2014).

\bibitem{ZhuJFM15} J.-Z. Zhu, Isotropic polarization of compressible flows.
    J. Fluid Mech., {\bf 787} 440 (2016)


\bibitem%[Celani et al.(2000)]
{CelaniETC00} A. Celani, A. Lanotte, A. Mazzino \& M. Vergassola,
Universality and saturation of intermittency in passive scalar turbulence.
Phys. Rev. Lett. {\bf 84}, 2385--2388 (2000).

\bibitem%[Bourouiba(2008a)]
{BourouibaPoF08} L. Bourouiba, Model of a truncated fast rotating flow at
infinite Reynolds number. {\em Phys. Fluid\/} {\bf 20}, 07512 (2008).

\bibitem%[Yamazaki, Kaneda \& Rubinstein(2002)]
{YKR02} Y. Yamazaki, Y. Kaneda \& R. Rubinstein, Dynamics of inviscid
truncated model of rotating turbulence. J. Phys. Soc. Jpn. {\bf 71}, 81
(2002).



\bibitem{footnotePhysicalMeasure}
%Some caveats: As said, the calculations do not directly exploint the information of the passiveness of the scalar, so one may put $\Gamma_{\mathcal{C}}=0$ in the expressions for $U_h$ and $W$, which however does not matter but is just a matter of re-transformation and re-interpretation of the parameters. And, Eq. (\ref{eq:2D3CU}) could make more sense for some relevant realistic situations such as the rotating flows, where the partial coupling makes the scalar not a purely passive one, to be discussed below. Of couse, we have assumed the applicability of the Gibbs measure, which implicitely involves issues such as ergodicity (or sufficient mixing) and equivalence of (microcanonical and canonical) ensembles, which actually are subtle and are not always granted, see, e.g., G. L. Eyink and H. Spohn, Negative-temperature states and large-scale, long-lived vortices in two-dimensional turbulence. J. Stat. Phys. \textbf{70}, 833 - 886 (1993). Concerning the ``physical measure'', discussed in
J.-P. Eckmann and D. Ruelle, Ergodic theory of chaos and strange attractors. Rev. Mod. Phys. {\bf 57}, 617 - 656 (1985). %, for the ideal Galerkin-truncated 2D3C system, it is indicated here that a measure, among the different invariant ones, is ``physical'' or not may depend on which physical quantities/observables are addressed and of course on which physical situations. The appearance of $\Gamma_{\mathcal{C}}$ in the statistics of the advecting field as superficially an unexpected back reaction, a paradox, nevertheless warns more general subtleties, but even if the formally calculated absolute equilibria were not numerically realizable/`physical' the indication from the result for the turbulent system could be made meaningful by the assistance of external ingredients. Even more fundamentally, as mentioned in the introductory discussion, statistical/probabilistic description by itself is inadequate for dynamical systems, in absolute-equilibrium or turbulent state, in general: Here, due to dynamical dependence of $\theta$ on $\zeta$, it is possible to have covariance between them, meaning statistically, but not dynamically, \textit{linear} dependence of $\zeta$ on $\theta$; also, from the joint Gaussian distribution of the applied Gibbs ensemble for the absolute equilibrium, zero covariance would indicate statistical, though not necessarily dynamical, independence between them.
The caveat is that the (Gibbs) measure may be ergodic for the projections
onto the $\theta$ space and onto the $\bm{v}$ space, but not in the
manifold, constrained by $\mathcal{C}$ and other invariants, in the whole
phase $\theta$-$\bm{v}$ space: Such partial-ergodicity presents constraint
on $\theta$ but not on $\bm{v}$. Note also that as pointed out By G. L.
Eyink and H. Spohn [``Negative-temperature states and large-scale,
long-lived vortices in two-dimensional turbulence.'' J. Stat. Phys.
\textbf{70}, 833 - 886 (1993)], strict ergodicity is not as easily satisfied
but is neither necessary condition to justify equilibrium statistics.


\bibitem{EyinkEuler08} G. L. Eyink, Dissipative anomalies in singular Euler
    flows. Phys. D \textbf{237}, 1956-1968 (2008).

\bibitem{SulemFrischJFM75} P. L. Sulem and U. Frisch, Bounds on energy flux
    for finite energy turbulence. J. Fluid Mech. \textbf{72}, 417-423
    (1975).


\bibitem{CMMVprl02} A. Celani, T. Matsumoto, A. Mazzino and M. Vergassola,
    Scaling and universality in turbulent convection. Phys. Rev. Lett. 88
    054503 (2002).

\bibitem%[Celani et al.(2002)]
{CCMVprl02} A. Celani, M. Cencini, A. Mazzino and M. Vergassola, Active vs
passive scalar turbulence. Phys. Rev. Lett. {\bf 89} 234502 (2002).

%\bibitem{CCGPepl02}
%E. S. C. Ching, Y. Cohen, T. Gilbert and I. Procaccia, Statistically preserved structures and anomalous scaling
%in turbulent active scalar advection. Europhys. Lett. 60 369 (2002).
%
%\bibitem{CCGPpre03}
%E. S. C. Ching, Y. Cohen, T. Gilbert and I. Procaccia, Active and passive fields in turbulent transport: The role of statistically preserved structures. Phys. Rev. E 67 016304 (2003).

\bibitem{Novikov} E.A. Novikov, Functionals and the method of random forces
    in turbulence theory, {\it Zh. Exper. Teor. Fiz.} {\bf 47}, 1919-1926
    (1964).

\bibitem%[Eyink et al. (2013)]
{EyinkNature13} G. Eyink, E. Vishniac, C. Lalescu, H. Aluie, K. Kanov, K.
B\"urger, R. Burns, C. Meneveau and A. Szalay, Flux-freezing breakdown in
high-conductivity magnetohydrodynamic turbulence. Nature {\bf 497}, 466
(2013).

\bibitem{FrischVillone14} U. Frisch and B. Villone, Cauchy's almost
    forgotten Lagrangian formulation of the Euler equation for 3D
    incompressible flow. arXiv:1402.4957v3.

\bibitem{EyinkJMP09} G. L. Eyink, Stochastic line motion and stochastic flux
    conservation for nonideal hydromagnetic models. J. Math. Phys.
    \textbf{50}, 083102 (2009).

\bibitem{FrischWirthLNP97} U. Frisch and A. Wirth, Intermittency of passive
    scalars in delta-correlated flow: Introduction to recent work. Lecture
    Notes in Physics \textbf{491}, 53 -- 64 (1997).

\bibitem{KramerMajdaVanden-Eijnden03} P. R. Kramer, A. Majda and E.
    Vanden-Eijden, Closure Approximations for Passive Scalar Turbulence: A
    Comparative Study on an Exactly Solvable Model with Complex Features.
    Journal of Statistical Physics \textbf{111}, 565 (2003).

\bibitem%[Waleffe(1992)]
{W92} F. Waleffe, The nature of triad interactions in homogeneous
turbulence. {\em Phys. Fluids A\/} {\bf 4}, 350--363 (1992).

%\bibitem{FluxFootnote}
%Intuitively, the stronger the concentration of the equilibrium spectrum at an end of the wavenumber interval is the stronger indication of a corresponding genuin turbulence flux should be; but how strong the concentration should be to be related to a non-zero flux, if indeed, seems completely clueless.

%\bibitem{GarnierAlemanySulemPouquetJdM81}
%M. Garnier, A. Alemany, P. L. Sulem and A. Pouquet, Influence of an external magnetic field on large scale low magnetic Reynolds number MHD turbulence. Journal de M\'ecanique \textbf{24}, 233 (1981).

\bibitem{CambonGodeferd93} C. Cambon and F. S. Godeferd, Inertial transfers
    in freely decaying rotating, stably stratified, and MHD turbulence. In
    Progress in Turbulence Research: Progress in Astronautics and
    Aeronautics (ed. H. Branover \& Y. Unger), AIAA \textbf{162}, 150-168
    (1993).

%\bibitem{MarionBertoglioCambonMathieuCRAS88}
%MarionBertoglioCambonMathieuCRAS88
%
%
%\bibitem{BatailleBertoglioMarionCRAS92}
%BatailleBertoglioMarionCRAS92

\bibitem%[Waleffe(1993)]
{W93} F. Waleffe, Inertial transfers in the helical decomposition. {\em
Phys. Fluids A\/} {\bf 5}, 677--685 (1993).




\bibitem%[Cencini, Muratore-Ginanneschi \& Vulpiani(2011)]
{CenciniMGV11} M. Cencini, P. Muratore-Ginaneschi \& A. Vulpiani, Nonlinear
Superposition of Direct and Inverse Cascades in Two-Dimensional Turbulence
Forced at Large and Small Scales. Phys. Rev. Lett. {\bf 107}, 174502 (2011).

\bibitem{FrischPRL08} U. Frisch, S. Kurien, R. Pandit, W. Pauls, S.S. Ray,
    A. Wirth, and J.-Z. Zhu, Galerkin Truncation, Hyperviscosity and
    Bottleneck in Turbulence. Phys. Rev. Lett. \textbf{101}, 144501 (2008).

\bibitem{ZhuTaylor10} J.-Z. Zhu and M. Taylor, Intermittency and
    Thermalization in Turbulence. Chin. Phys. Lett. \textbf{27}, 054702
    (2010).


%\bibitem{GaltierPRE03}
%S. Galtier, Weak inertial-wave turbulence theory. Phys. Rev. E \textbf{68}, 015301(R) (2003).
%
%\bibitem{BelletCambonJFM06}
%F. Bellet, F.S. Godeferd, and F.S. Scott, Wave turbulence in rapidly rotating flows. J. Fluid Mech.
%\textbf{562}, 83 (2006).
%
%\bibitem{ScottJFM14}
%J. F. Scott, Wave turbulence in a rotating channel. J. Fluid Mech. \textbf{741}, 316-349 (2014).
%
%\bibitem{HideGFD76}
%R. Hide, A note on helicity. Geophys. Astrophys. Fluid Dyn. \textbf{7}, 157-161 (1976)
%
%\bibitem{MoffattBook}
%K. Moffatt, Generation of magnetic fields in conducting fluids. Cambridge University Press (1978).
%
%\bibitem{Riley}
%J. J. Riley, and S. M. de Bruyn Kops, Dynamics of turbulence strongly infuenced by buoyancy. Phys. Fluids \textbf{15}, 2047-2059(2003).
%
%\bibitem{Bartello95}
%P. Bartello, Geostrophic adjustment and inverse cascades in rotating stratified turbulence. J. Atmos. Sci. \textbf{52}, 4410-4428 (1995).
%
%\bibitem{SagautCambon08}
%P. Sagaut and C. Cambon, Homogeneous turbulence dynamics. Cambridge University Press (2008).
%
%\bibitem{RoraiPRE13}
%C. Rorai, D. Rosenberg, A. Pouquet and P. D. Mininni, Helicity dynamics in stratified turbulence in the absence of forcing. Phys. Rev. E \textbf{87}, 063007 (2013).
%
%\bibitem{AluiKurien11}
%H. Aluie and S. Kurien, Joint downscale fluxes of energy and potential enstrophy in rotating stratified Boussinesq flows. Europhys. Lett. \textbf{96} (4), 44006 (2011).
%
%\bibitem{Wingate12}
%B. A. Wingate, P. Embid, M. Holmes-Cerfon and M. A. Taylor, Low Rossby limiting dynamics for stably stratified flow with finite Froude number. J. Fluid Mech. \textbf{676}, 546 (2011).
%
%\bibitem{PouquetMarino13}
%A. Pouquet and R. Marino, Geophysical turbulence and the duality of the energy flow across scales. Phys. Rev. Lett. \textbf{111}, 234501 (2013).

\bibitem{Kain08} J. S. Kain, et al., Some Practical Considerations Regarding
    Horizontal Resolution in the First Generation of Operational
    Convection-Allowing NWP. Weather and Forcasting \textbf{23}, 931 (2008).

\bibitem%[Montgomery \& Turner(1982)]
{MontgomeryTurnerPoF82} D. Montgomery \& L. Turner,
Two-and-a-half-dimensional magnetohydrodynamic turbulence. Phys. Fluids {\bf
25}, 345--349 (1982).

\bibitem%[Biferale, Musacchio \& Toschi(2012)]
{bmt12} L. Biferale, S. Musacchio and F. Toschi, Inverse energy cascade in
three-dimensional isotropic turbulence. {\em Phys. Rev. Lett.\/} {\bf 108}
104501 (2012).

%\bibitem{BiferaleTitiJST13}
%L. Biferale and E. Titi, On the Global Regularity of a Helical-Decimated Version
%of the 3D Navier-Stokes Equations. J. Stat. Phys. {\bf 151}, 1089 (2013).

\bibitem{footnote} It may make it clearer to quote from Ref. \cite{CCEH05}
    that ``the present theorems effectively say nothing about the validity
    of the resonant wave theory at hight $Re$, for realistic values of the
    Rossby number'' and ``the asymptotic analysis does not determine the
    largest possible time'' for which theory is proved to be valid.



\bibitem{BrandenburgETC} A. Brandenburg, W. Dobler \& K. Submanian. Magnetic
    helicity in stellar dynamos: new numerical experiments. Astron. Nachr.
    \textbf{323}, 99 - 123 (2002).

\bibitem{ZhuHammettPoF10} J.-Z. Zhu and G. Hammett, Gyrokinetic absolute
    equilibrium and turbulence. Phys. Plasmas \textbf{17}, 122307 - 1 - 13
    (2010).


\bibitem{footnoteLevichTzvetkov84} E. Levich and E. Tzvetkov, Helical
    cyclogenesis, Phys. Rep. {\bf 100A}, 53 (1984).

\bibitem{LevinaMontgomery14} G. V. Levina and M. T. Montgomery, Numerical
    Diagnosis of Tropical Cycolgenesis Based on a Hypothesis of Helical
    Self-Organization of Moist Convective Atmospheric Turbulence. Doklady
    Earth Sciences {\bf 458}, 1143 (2014).

\bibitem{ConstantinMajdaCMP88} P. Constantin and A. Majda, The Beltrami
    Spectrum for Incompressible Fluid Flows. Commun. Math. Phys. {\bf 115},
    435 (1988).

\bibitem{GallavottiCohen95} G. Gallavotti and E. G. D. Cohen, Dynamical
    ensembles in nonequilibrium statistical mechanics. Phys.  Rev.  Lett.
    {\bf 74}, 2694 (1995).

%\bibitem{ZhuCompressible15}
%J.-Z. Zhu, Isotropic polarization of compressible flows. J. Fluid Mech. {\bf 787}, 440 (2016).
%
%\bibitem%[Kraichnan(1955)]
%{k55}
%R. H. Kraichnan, On the Statistical Mechanics of an Adiabatically Compressible Fluid.
%{\em The Journal of the Acoustical Society of America \/} {\bf 27}, 438--441 (1955).

\bibitem%[Moses(1971)]
{Moses71} H. E. Moses, Eigenfunctions of the curl operator, rotationally
invariant Helmholtz theorem and applications to electromagnetic theory and
fluid mechanics. SIAM ~(Soc. Ind. Appl. Math.) J. Appl. Math. {\bf 21}, 114
(1971).


\end{thebibliography}
\end{document}